\documentclass[draft]{agujournal}
\usepackage{apacite}
\usepackage{url} 

\journalname{JGR-Space Physics}

\usepackage{graphicx}

\usepackage{amsmath}
\usepackage{amssymb}

\usepackage{ccaption}             

\usepackage[shortlabels]{enumitem}

\begin{document}

%
%


\title{Inferring source properties of monoenergetic electron precipitation from
  kappa and Maxwellian moment-voltage relationships}

%
%




\authors{Spencer M. Hatch,\affil{1}\thanks{Department of Physics and Technology,
    All\'{e}gaten 55, N-5007 Bergen, Norway}
  James LaBelle\affil{2},
  Christopher C. Chaston\affil{3}
  }

\affiliation{1}{Birkeland Center for Space Science, University of Bergen,
  Bergen, Norway}
\affiliation{2}{Department of Physics and Astronomy, Dartmouth College, Hanover,
  New Hampshire, USA.}
\affiliation{3}{Space Sciences Laboratory, University of California, Berkeley,
  California, USA}


\correspondingauthor{S. M. Hatch}{Spencer.Hatch@uib.no}



\begin{keypoints}
\item The magnetospheric source parameters that account for observed auroral
  electron distributions are derived from a generalized kinetic model.
\item The degree of non-thermality (kappa index) in auroral primaries is
  required to correctly prescribe the magnetospheric source parameters.
\item We present the first analytical kinetic model for the relationship between
  density and acceleration potential along auroral field lines.
\end{keypoints}

%
%

\begin{abstract}
  
  We present two case studies of FAST electrostatic analyzer measurements of
  both highly nonthermal ($\kappa \lesssim$~2.5) and weakly nonthermal/thermal
  monoenergetic electron precipitation at $\sim$4000~km, from which we infer the
  properties of the magnetospheric source distributions via comparison of
  experimentally determined number density--, current density--, and energy
  flux--voltage relationships with corresponding theoretical relationships. We
  also discuss the properties of the two new theoretical number density--voltage
  relationships that we employ. Moment uncertainties, which are calculated
  analytically via application of the \citet{Gershman2015} moment uncertainty
  framework, are used in Monte Carlo simulations to infer ranges of
  magnetospheric source population densities, temperatures, $\kappa$ values, and
  altitudes. We identify the most likely ranges of source parameters by
  requiring that the range of $\kappa$ values inferred from fitting experimental
  moment-voltage relationships correspond to the range of $\kappa$ values
  inferred from directly fitting observed electron distributions with
  two-dimensional kappa distribution functions. Observations in the first case
  study, which are made over $\sim$78--79$^\circ$ invariant latitude (ILAT) in
  the Northern Hemisphere and 4.5--5.5 magnetic local time (MLT), are consistent
  with a magnetospheric source population density $n_m =$~0.7--0.8~cm$^{-3}$,
  source temperature $T_m \approx$~70~eV, source altitude $h =$~6.4--7.7~$R_E$,
  and $\kappa =$~2.2--2.8. Observations in the second case study, which are made
  over 76--79$^\circ$~ILAT in the Southern Hemisphere and $\sim$21~MLT, are
  consistent with a magnetospheric source population density
  $n_m =$~0.07--0.09~cm$^{-3}$, source temperature $T_m \approx$~95~eV, source
  altitude $h \gtrsim$~6~$R_E$, and $\kappa =$~2--6.

\end{abstract}

%
%
%


%
%

  \section{Introduction}

  Potential differences exist along geomagnetic field lines that connect the
  plasma sheet and high-latitude magnetosphere to the
  ionosphere. \citet{Knight1973} formally demonstrated the relationship between
  a field-aligned, monotonic potential profile represented by a total potential
  difference $\Delta \Phi$ and field-aligned current density below the potential
  drop $j_{\parallel}$ generated by precipitation of magnetospheric electrons
  subject to a magnetic mirror ratio $R_B = B/B_{m}$,
  \begin{linenomath*}
    \begin{equation} \label{eqKnight} j_{\parallel,M} ( \, \Delta \Phi ; T_m,
      n_m, R_B ) = - e n_m \left ( \frac{T_m}{2 \pi m_e} \right )^{\frac{1}{2}}
      R_B \left [ 1 - \left ( 1 - R_B^{-1} \right ) \textrm{exp} \left \{ -
      \frac{\overline{\phi}}{( R_B - 1 )} \right \} \right].
  \end{equation}
  \end{linenomath*}
  Here $T_m$ and $n_m$ are the temperature and density of precipitating
  electrons at the magnetospheric source,
  $\overline{\phi} \equiv e \Delta \Phi / T_m$ is the potential drop normalized
  by source temperature, and $m_e$ is the electron mass. The subscript $m$
  indicates the magnetospheric source region.

  The J-V relation (\ref{eqKnight}) assumes that the magnetospheric source
  population is isotropic and in thermal equilibrium, and is thus described by a
  Maxwellian distribution. However, magnetospheric electron and ion
  distributions observed with spacecraft often show suprathermal tails
  \citep{Christon1989,Christon1991,Wing1998,Kletzing2003} which may be produced
  via a number of mechanisms (See e.g., review by \citealp{Pierrard2010}.) The
  possibility of a source distribution with a ``high energy tail'' was in fact
  acknowledged by \citet{Knight1973}, and reformulations of the Maxwellian J-V
  relation (\ref{eqKnight}) assuming a variety of alternative source
  distributions have been developed
  \citep{Pierrard1996,Janhunen1998,Dors1999,Bostrom2003a,Bostrom2004}. One such
  alternative distribution employed with increasing frequency is the isotropic
  kappa distribution
  \begin{linenomath*}
    \begin{equation} \label{eqKappa1D} f_{\kappa}(E; T, n, \kappa) = n \left (
      \frac{m}{2 \pi T \, (1 - \frac{3}{2 \kappa}) } \right )^{\frac{3}{2}}
      \frac{\Gamma \left ( \kappa + 1 \right ) }{ \kappa^{3/2} \Gamma \left ( \kappa -
        \frac{1}{2} \right ) } \left ( 1 + \frac{E}{ \left ( \kappa - \frac{3}{2}
        \right ) T } \right )^{-1 - \kappa},
    \end{equation}
  \end{linenomath*}
  which originally entered the space physics community as a model for
  high-energy tails of observed solar wind plasmas \citep{Vasyliunas1968}. The
  additional parameter $\kappa \in [ \, \kappa_{\textrm{min}}, \infty )$
  parameterizes the degree
  \begin{equation}
    \label{eqCorr}
    \rho = \kappa_{\textrm{min}}/\kappa \in \left (0,1 \right ] 
  \end{equation}
  to which particle motion is correlated, and is related to the ``thermodynamic
  distance'' between a stationary (i.e., invariant over relevant time scales)
  non-equilibrium state and thermal equilibrium. The range of $\kappa$ values
  observed in a plasma environment depends on the transport, wave-particle
  interaction, and acceleration processes that are found within that environment
  \citep[see, e.g.,][]{Treumann1999, Pierrard2010}. The theoretical minimum (in
  three dimensions) $\kappa_{\textrm{min}} = 3/2$ corresponds to perfectly
  correlated degrees of freedom and particle motions ($\rho = 1$), while
  $\kappa \rightarrow \infty$ corresponds to uncorrelated degrees of freedom
  ($\rho = 0$) and thermal equilibrium, or a Maxwellian distribution
  \citep{Treumann1999a}.

  \citet{Livadiotis2010} have shown that $\kappa_t \simeq 2.45$
  ($\rho \simeq 0.61$) marks a transition between these two extremes, with
  $ \kappa_{\textrm{min}} \leq \kappa \lesssim \kappa_t$ constituting the
  ``far-equilibrium'' regime, and $ \kappa_t \lesssim \kappa < \infty$ the
  ``near-equilibrium'' regime.

  Relaxing the assumption of a magnetospheric source population in thermal
  equilibrium, \citet{Dors1999} showed that the J-V relation (\ref{eqKnight})
  becomes
  \begin{linenomath*}
    \begin{align}
      \label{eqDors}
      \begin{array}{ll}
        j_{\parallel,\kappa} ( \, \Delta \Phi ; T_m, n_m, \kappa, R_B ) &= - e n_m \left ( \frac{ (1 - \frac{3}{2 \kappa}) \, T_m}{ 2 \, \pi \, m_e} \right)^{1/2} \frac{\Gamma (\kappa + 1)}{ \kappa^{3/2} \Gamma (\kappa - \frac{1}{2})} \frac{R_B}{1 - 1/\kappa} \\
       &\times \left[1 - \left(1 - R_B^{-1} \right) \left(1+\frac{\overline{\phi}}{ (\kappa - \frac{3}{2})(R_B - 1)} \right)^{1-\kappa} \right]. \\
      \end{array}
    \end{align}
  \end{linenomath*}
  For equal $n_m$ and $T_m$, the values of $j_\parallel$ predicted by equation
  (\ref{eqKnight}) and equation (\ref{eqDors}) differ by more than $\sim$33\%
  for the ``far-equilibrium'' regime
  ($\kappa_{\textrm{min}} < \kappa \lesssim \kappa_t$) \citep{Hatch2018}. Recent
  case studies \citep{Kaeppler2014a,Ogasawara2017} and a statistical study
  \citep{Hatch2018} suggest that such extreme $\kappa$ values seldom occur in
  the auroral acceleration region.

  Equation (\ref{eqKnight}),otherwise known as the Knight relation, and Equation
  (\ref{eqDors}) are examples of current density--voltage (J-V)
  relationships. Such relationships are a means for understanding the role of
  field-aligned potential differences within large-scale magnetospheric current
  systems. The Knight relation in particular has contributed to present
  understanding of the magnetosphere-ionosphere current system
  \citep[e.g.,][]{Shiokawa1990,Lu1991,Temerin1997,Hultqvist1999,Cowley2000,Bostrom2003a,Paschmann2003,Pierrard2007a,Karlsson2012,Dombeck2013}.
  Moment-voltage relationships such as that between energy flux and voltage, a
  ``J$_E$-V'' relationship, are also derivable
  \citep{Chiu1978,Pierrard1996,Janhunen1998,Liemohn1998,Dors1999,Bostrom2003a,Bostrom2004,Pierrard2007a}. These
  other relationships have received comparatively little attention even though
  they represent additional, valuable tools for estimating magnetospheric source
  population parameters from particle observations at lower altitudes.

  Using previously published J-V and J$_E$-V relationships and two new number
  density--voltage (n-V) relationships, we show how knowledge of the degree to
  which monoenergetic precipitation departs from Maxwellian form leads
  to identification of narrow ranges of magnetospheric source parameters that are
  compatible with observed moment-voltage relationships. This technique is
  enabled by the \citet{Gershman2015} methodology for analytic calculation of
  moment uncertainties as well as direct two-dimensional distribution fits.

  \section{Methodology}

  Here we summarize the J$_E$-V and n-V relationships that we use in addition to
  the J-V relationships (\ref{eqKnight}) and (\ref{eqDors}), as well as the
  \citet{Gershman2015} methodology for estimating moment uncertainties of
  measured electron distribution functions.

  \subsection{J$_E$-V and n-V relationships}

  Assuming a monotonic potential profile the energy flux--voltage (J$_E$-V)
  relationships for isotropic Maxwellian and kappa source distributions are
  respectively \citep{Dors1999}
  \begin{linenomath*}
    \begin{equation}\label{eqJEM}
      j_{E\parallel,M} = \frac{n_m T^{3/2}}{\sqrt{2 \pi m_e}}R_B \left\{ (2+\overline{\phi}) - \left[\overline{\phi}+2\left(1-R_B^{-1}\right)\right] \exp \left(\frac{-\overline{\phi}}{R_B-1}\right)\right\}
    \end{equation}
  \end{linenomath*}
  and 
  \begin{linenomath*}
    \begin{align}
      \label{eqJEK}
      \begin{array}{ll}
        j_{E\parallel,\kappa}  = & n_m \frac{ \left[(1 - \frac{3}{2 \kappa}) \, T_m \right]^{3/2}}{ \left (2 \, \pi \, m_e \right)^{1/2}} \frac{\kappa^2}{ \kappa^{3/2} \left(\kappa-1\right) \left(\kappa-2\right)} \frac{\Gamma \left(\kappa + 1\right) }{ \Gamma \left(\kappa - \frac{1}{2}\right)} R_B \Bigg \{ \left[2+\frac{\kappa - 2}{\kappa-3/2} \overline{\phi} \right] \\ 
        &- \Pi^{-\kappa+1} \left( 1 - {R_B}^{-1} \right ) \, \, \left ( \frac{ \kappa-2 }{ \kappa - 1 } + \frac{\kappa-2}{\kappa-3/2} \overline{\phi} \right ) \left ( \frac{\kappa}{ \left ( \kappa - 1 \right ) \left(R_B - 1 \right ) } + 1 \right ) \\
        &- \Pi^{-\kappa+2} \left(1-{R_B}^{-1} \right)^2 \left( 1 + \frac{ 1 + \kappa / \left( R_B - 1 \right) }{\kappa - 1} \right) \Bigg \},
      \end{array}
    \end{align}
  \end{linenomath*}
  with $\Pi = 1 + \frac{\overline{\phi}}{(\kappa-3/2)(R_B-1)}$ in
  (\ref{eqJEK}). One may also derive the corresponding Maxwellian and kappa n-V
  relationships (Appendix A) 
  \begin{equation}\label{eqDensVM}
    n_M ( \Delta \Phi ; T_m,
    n_m, R_B ) = n_m \left[\frac{1}{2} e^{\overline{\phi}} \textrm{erfc} \left( \sqrt{\overline{\phi}} \right) + \sqrt{\frac{R_B-1}{\pi}} D \left(\sqrt{\frac{\overline{\phi}}{R_B-1}}\right)\right];
  \end{equation}
  \begin{equation}\label{eqDensVK}
    \begin{aligned}
      n_\kappa( \Delta \Phi ; T_m,
      n_m, R_B, \kappa ) = &&& n_m \frac{\overline{\phi}^{3/2}}{\sqrt{\pi} (\kappa-\frac{3}{2})^{3/2}} \frac{\Gamma(\kappa+1)}{\Gamma(\kappa-1/2)} \\
      & \times && \Bigg[ \int_{0}^{\infty} d x \sqrt{1+x} \left(1+\frac{x \overline{\phi}}{\kappa-\frac{3}{2}} \right)^{-\kappa-1} \\
        &&& - \frac{1}{R_B-1} \int_{0}^{1} dx \sqrt{1-x} \left(1+\frac{x
            \overline{\phi}}{\left(\kappa-\frac{3}{2}\right)\left(R_B-1\right)}
        \right)^{-\kappa-1} \Bigg];
    \end{aligned}
  \end{equation}
  where $D(y) = \exp\left(-y^2\right) \int_0^y \exp\left(y'^2\right) dy'$ in
  (\ref{eqDensVM}), and $x$ is a dummy integration variable in
  (\ref{eqDensVK}). The relationships (\ref{eqJEM})--(\ref{eqDensVK}) are
  written in terms of the same variables as those used in the J-V relationships
  (\ref{eqKnight}) and (\ref{eqDors}). Properties of the J$_E$-V relationships
  (\ref{eqJEM}) and (\ref{eqJEK}) have been discussed by \citet{Dors1999}. Some
  properties of the n-V relationships (\ref{eqDensVM}) and (\ref{eqDensVK}),
  which are previously unpublished, are discussed here.

  \begin{figure}
    \centering
    \noindent\includegraphics[width=1.0\textwidth]{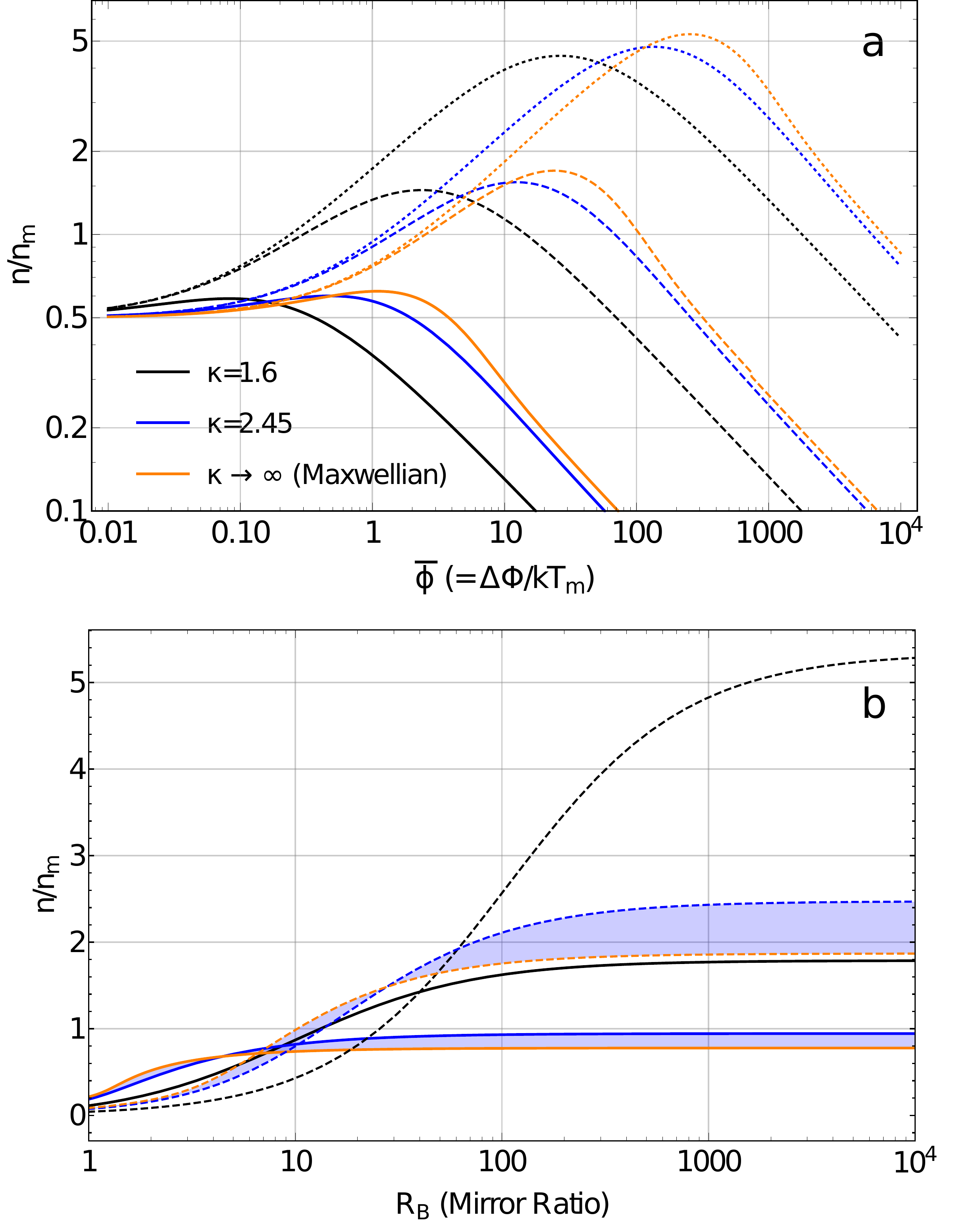}
    \caption{Number density--voltage (n-V) relationships in Equations
      (\ref{eqDensVK}) and (\ref{eqDensVM}). The ratio $n/n_m$ is plotted on the
      $y$ axis, where $n_m$ is magnetospheric source population density, and $n$
      is the density at the altitude corresponding to mirror ratio $R_B$ and to
      the bottom of normalized potential drop
      $\overline{\phi} = \Delta \Phi / T_m$. (a) n-V relationships as a function
      of $\overline{\phi}$ for $R_B =$~3 (solid lines), $R_B =$~30 (dashed
      lines), and $R_B =$~300 (dotted lines). (b) n-V relationships as a
      function of $R_B$ for $\overline{\phi} =$~1 (solid lines) and
      $\overline{\phi} =$~10 (dashed lines). In Figure 1b the region between the
      $\kappa = \kappa_t$ and $\kappa \rightarrow \infty$ curves is shaded. }
    \label{FigDens}
  \end{figure}

  Figure~\ref{FigDens}a shows the n-V relationships as a function of
  $\overline{\phi}$ for Maxwellian ($\kappa \rightarrow \infty$, orange lines),
  moderately nonthermal ($\kappa = \kappa_t$, blue lines), and extremely
  nonthermal ($\kappa = 1.6$, blue lines) source populations. Mirror ratios of
  3, 30, and 300 are respectively represented by solid, dashed, and dotted
  lines. 

  It is evident that $n / n_m \rightarrow 1/2$ in the limit
  $\overline{\phi} \rightarrow$~0; this behavior is shown analytically for the
  Maxwellian n-V relation in the asymptotic expression
  (\ref{eqDensVMlittlePhi}). On the other hand $n / n_m \rightarrow 0$ for
  $\overline{\phi} \gg R_B$, as shown in the asymptotic expression
  (\ref{eqDensVMlittleRB}). 

  The n-V relationships predict half the source density $n_m$ in the
  $\overline{\phi} \rightarrow$~0 limit (i.e., no field-aligned potential)
  because only those particles in the magnetospheric source region having a
  parallel velocity component toward the ionosphere (defined as
  $v_\parallel > 0$) are included in the range of integration used to obtain the
  n-V relationships; all others move away from the ionosphere and are ignored
  (Appendix A). This restriction on the range of integration is identically the
  reason that in the $\overline{\phi} \rightarrow$~0 limit the J-V relations
  (\ref{eqKnight}) and (\ref{eqDors}) and the J$_E$-V relationships
  (\ref{eqJEM})--(\ref{eqJEK}) respectively predict nonzero current densities,
  or ``thermal flows'' \citep{Paschmann2003}, and nonzero energy fluxes. For
  instance, in this limit $j_{\parallel,M} = -e n_m (T_m / 2 \pi
  m_e)^{1/2}$. (See, e.g., Figure 3.8 in \citealp{Paschmann2003}.)

  More generally, increasing $\overline{\phi}$ (i) increases the number of
  particles that have access to lower altitudes by increasing $v_\parallel$,
  which increases $n/n_m$ (the total volume under the distribution function),
  and (ii) compresses the distribution function in velocity space (see
  Inequality (\ref{eqWCondition})), which decreases $n/n_m$. On the other hand
  increasing $R_B$ while simultaneously conserving the first adiabatic invariant
  causes the distribution function to evolve toward an annular or ``ring''
  distribution, such that (i) particles at large pitch angles
  ($v_{\perp} > v_\parallel$) in the source region are reflected at lower
  altitudes, which decreases $n/n_m$, and (ii) particles with small pitch angles
  ($v_{\parallel} > v_\perp$) in the source region, particularly those near the
  peak of the distribution near $v_\parallel = \sqrt{2 e \Delta \Phi / m}$ also
  spread to larger pitch angles, increasing the total volume under the
  distribution function and thereby increasing $n/n_m$. The maximum values of
  $n/n_m$ in Figure~\ref{FigDens}a thus represent nontrivial interplay of these
  competing factors (see Equation~\ref{EqMaxValWRTPhi}).

  Figure~\ref{FigDens}b shows as a function of $R_B$ the Maxwellian and kappa
  n-V relationships (\ref{eqDensVM}) and (\ref{eqDensVK}) normalized by source
  density $n_m$, for $\overline{\phi} =$~1 (solid lines), $\overline{\phi} =$~10
  (dashed lines), and several values of $\kappa$. The magnetospheric source
  region corresponds to $R_B = 1$. For a Maxwellian source population in the
  limit $R_B \rightarrow 1$, $n$ is given by the first term in the Maxwellian
  n-V relation (\ref{eqDensVK}). For a kappa source population in the limit
  $R_B \rightarrow 1$, $n$ approaches values similar to that approached by the
  Maxwellian curve. The topmost curve in Figure~\ref{FigDens}b shows that for
  $\overline{\phi} = 10 \ll R_B$ and $\kappa = 1.6$, the density at lower
  altitudes increases by as much as a factor 5. More generally,
  $n/n_m \propto \sqrt{\overline{\phi}/(1-\frac{3}{2 \kappa})}$ for
  $1 \lesssim \overline{\phi} \ll R_B$. The shaded region between the
  $\kappa = \kappa_t$ and $\kappa \rightarrow \infty$ curves (blue and orange,
  respectively) in Figure~\ref{FigDens}b indicates that there is little
  difference, generally less than 30\%, between the Maxwellian and kappa n-V
  relationships for $\kappa \gtrsim \kappa_t$ and equal
  $\overline{\phi}$. Asymptotic expressions for the Maxwellian n-V relation
  (\ref{eqDensVM}) are given in Appendix~A.

  \subsection{Uncertainty of Distribution Moments}

  This study also relies on moments of measured electron distributions,
  including the number density $n$, field-aligned current density
  $j_\parallel = e \langle n v_\parallel \rangle$ and field-aligned energy flux
  $j_{E\parallel} = \frac{m}{2} \langle n v^2 v_\parallel \rangle$.  Estimation
  of the uncertainty of moments has typically involved generation of statistics
  of each moment via Monte Carlo simulation of $f(\mathbf{v})$
  \citep[e.g.,][]{Moore1998}. We alternatively use standard techniques of
  linearized uncertainty analysis to derive analytic expressions for the
  uncertainties of field-aligned current density and energy flux, respectively
  $\sigma_{j_\parallel}$ and $\sigma_{j_{E\parallel}}$, as functions of moments
  of $f(\mathbf{v})$ and moment covariances. (The uncertainty of number density
  $n$ is trivially $\sigma_{n}$.) Moment covariances are calculated following
  the methodology of \citet{Gershman2015}. In Appendix B we present both these
  analytic expressions and a summary of the \citet{Gershman2015} methodology,
  which together enable the Monte Carlo simulations presented in
  sections~\ref{sAfterData} and \ref{sAfterData2}.

  \section{Orbit 1607}


  \begin{figure}
    \centering
    \noindent\includegraphics[width=0.75\textwidth]{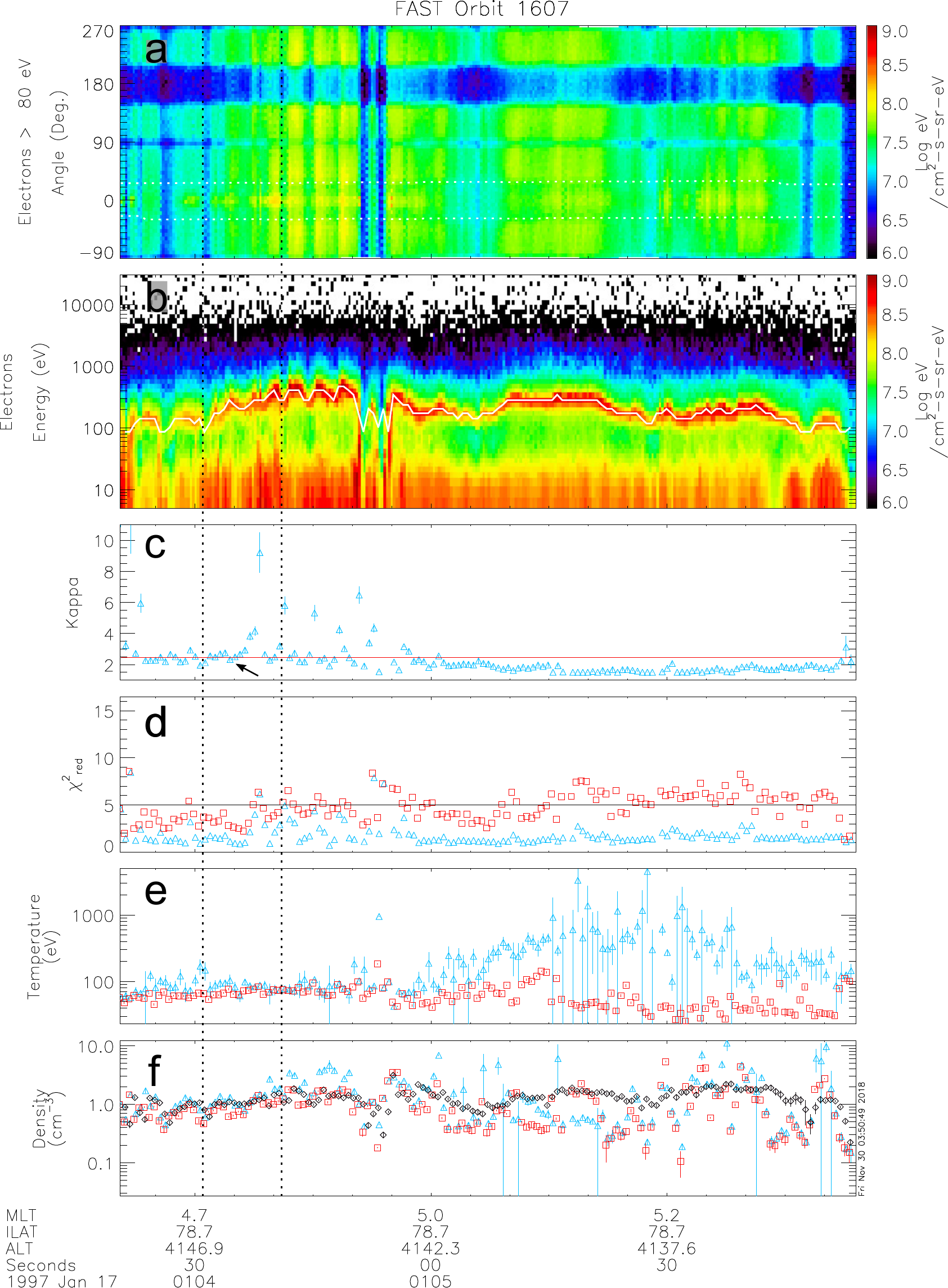}
    \caption{EESA observations of inverted V precipitation on Jan
      17, 1997, and corresponding 2-D fit parameters. Parameters related to
      best-fit Maxwellian and kappa distributions are respectively indicated by
      red squares and blue triangles in panels c--f. (a) $>$80~eV electron
      pitch-angle distribution. The earthward portion of the loss cone (see text
      for definition) comprises the $\sim$60$^\circ$ range of pitch angles
      between the dotted horizontal lines at $\sim$30$^\circ$ and
      $\sim$-30$^\circ$. (b) Average electron energy spectrum within the
      earthward loss cone. (c) $\kappa$ fit parameter for the best-fit kappa
      distribution. The red horizontal line indicates
      $\kappa = \kappa_t \simeq$~2.45. The black arrow indicates the kappa value
      for fits shown in Figure~\ref{Fig3}. (d) Reduced chi-squared statistic
      $\chi^2_{\mathrm{red}}$ for each fit type. The black horizontal line
      indicates $\chi^2_{\mathrm{red}} = 5$. (e) Best-fit temperatures. (f)
      Calculated density moments (black diamonds) and best-fit
      densities. Calculated densities are obtained as 2-D model-independent
      moments of the differential flux measured over pitch angles
      $\vert \theta \vert \leq$~30$^\circ$, from the energy of the channel
      immediately below the peak energy $E_p$ up to 5~keV. Resulting density
      moments and uncertainties (vertical black bars) are then multiplied by the
      ratio of the solid-angle ratio $1/(1-\cos 30^\circ) \approx 7.46$.
      Uncertainties of best-fit parameters represent 90\% confidence intervals
      obtained via Monte Carlo simulations with $N=$~5,000 trials. EESA
      observations are integrated to obtain an effective sample period
      $T = 0.63$~s.}
    \label{FigSum}
  \end{figure}


  \subsection{Data Presentation}

  During an approximately 90-s interval on Jan 17, 1997, the FAST satellite
  observed inverted V electron precipitation over 80--600~eV
  (Figures~\ref{FigSum}a and \ref{FigSum}b) and over $\sim$4.5--5.5 magnetic
  local time (MLT) in the Northern Hemisphere during low geomagnetic activity
  ($K_p =$~0$^-$). Figure~\ref{FigSum}a shows that over much of this interval
  the distributions include both isotropic and trapped components, while the
  anti-earthward loss cone is relatively depleted. For instance over
  01:05:10--01:05:15~UT the isotropic component is somewhat weak
  ($dJ_E/dE\lesssim 3\times 10^7$eV/cm$^2$-s-sr-eV) and the trapped component is
  relatively more intense ($dJ_E/dE\gtrsim 10^8$eV/cm$^2$-s-sr-eV).

  Figure~\ref{FigSum}b gives the observed electron energy spectrogram averaged
  over observations at all pitch angles within the earthward loss cone. The loss
  cone is calculated from model geomagnetic field magnitudes at FAST and at the
  100-km ionospheric footpoint, which are both obtained from International
  Geomagnetic Reference Field 11 (IGRF 11). For the period indicated between
  dashed lines (01:04:28--01:04:41~UT), which we will discuss momentarily,
  Figure~\ref{FigSum}b shows that the peak energy of monoenergetic electron
  precipitation varies between 80~eV and 500~eV.

  We perform full 2-D fits to the portion of electron distributions that are
  observed within the earthward loss cone (horizontal dotted white lines in
  Figure~\ref{FigSum}a) and between the energy at which the distributions peak
  above 80~eV ($E_p$) up to the 30-keV limit of FAST electron electrostatic
  analyzers (EESAs) \citep{Carlson2001}. To obtain these fits, we first form a
  1-D differential number flux distribution by averaging the counts within each
  EESA energy-angle bin over the range of angles within the earthward portion of
  the loss cone, after which 1-D fits of the resulting average differential
  number flux spectrum are performed using the model differential number flux
  $J = \frac{2 E}{m^2} f \left ( E- E_p \right)$, with
  $f \left ( E- E_p \right)$ either a 1-D Maxwellian or 1-D kappa
  distribution. The resulting 1-D best-fit parameters then serve as initial
  estimates for 2-D fits of the observed differential energy flux spectrum, over
  the previously described range of pitch angles and energies, using model
  differential energy flux
  $dJ_E/dE = \frac{2 E^2}{m^2} f \left ( E- E_p \right)$. Both 1-D and 2-D
  distribution fits are performed using Levenberg-Marquardt weighted
  least-squares minimization via the publicly available Interactive Data
  Language MPFIT library (http://cow.physics.wisc.edu/~craigm/idl/fitting.html).

  The most probable fit parameters and 90\% confidence intervals are then
  obtained by following a procedure similar to those recently employed by
  \citet{Kaeppler2014a} and \citet{Ogasawara2017}: for each time and each type
  of distribution, we fit $N =$~5,000 Monte Carlo simulated 2-D distributions by
  adding to each best-fit distribution a normal random number
  $Z \sim \mathcal{N}(0,1)$ that is multiplied by the counting uncertainty
  (section 15.6 in \citealp{Press2007}) in units of differential energy
  flux. For the simulated kappa distribution fits we also select a uniform
  random number $K \sim U(\kappa_{\textrm{min}},35)$ as an initial guess for
  $\kappa$.

  For each fit parameter we then form a histogram from the resulting 5,000 Monte
  Carlo values. The value at which the histogram peaks is taken to be the most
  probable fit parameter. We then use a simple algorithm that increases the size
  of a window centered on the most probable fit parameter until 4,500 (90\%) of
  the parameter values are included in the window.

  Reported parameters are $\kappa$ (Figure~\ref{FigSum}c) for the kappa
  distribution, and temperature (Figure~\ref{FigSum}e) and density
  (Figure~\ref{FigSum}f) for both Maxwellian and kappa distribution fits. For
  each most probable fit parameter the 90\% confidence interval is indicated by
  a vertical bar. In many instances the upper and lower limits of the 90\%
  confidence interval are very near the most probable fit parameter value. For
  example, for over 92\% of the most probable $\kappa$ parameters in
  Figure~\ref{FigSum}c the upper and lower limit of the 90\% confidence interval
  is within 10\% of the most probable $\kappa$ parameter itself.


  \begin{figure}
    \centering
    \noindent\includegraphics[width=1.0\textwidth]{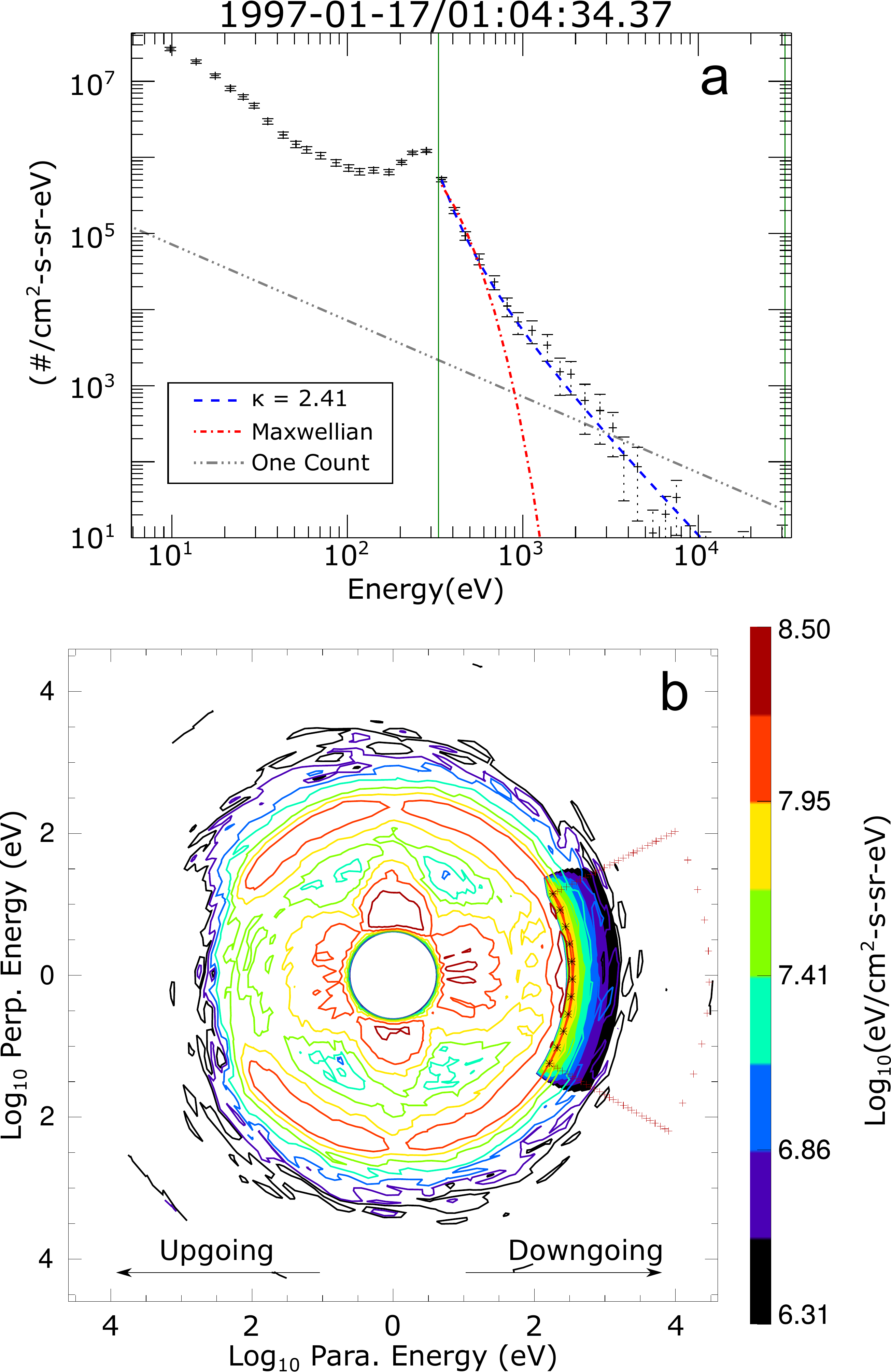}
    \caption{Electron spectra observed at 01:04:34.37--01:04:35.00~UT. (a) 1-D
      differential number flux spectrum (black crosses) obtained by averaging
      differential number flux spectra over all pitch angles within the
      earthward loss cone, with best-fit Maxwellian and kappa distributions
      overlaid (red dash-dotted line and blue dashed line, respectively). The
      uncertainty of each observed differential number flux is calculated by
      conversion of the electron count uncertainty $\sqrt{N}$ to units of
      differential number flux. (b) Best-fit 2-D kappa distribution (solid
      contours) with the observed 2-D differential energy flux spectrum overlaid
      (contour lines). The color bar at right shows the differential energy flux
      of each contour. For each pitch angle black asterisks indicate the peak
      energy $E_p = 315$~eV, and red plus signs outline the range of energies
      and pitch angles used to perform the 2-D fit.}
    \label{Fig3}
  \end{figure}

  %

  Figure~\ref{Fig3} shows example 1-D (Figure~\ref{Fig3}a) and 2-D
  (Figure~\ref{Fig3}b) distribution fits to electron observations during
  01:04:34.37--01:04:35.00~UT, indicated by the black arrow in
  Figure~\ref{FigSum}c. The observed distribution in Figure\ref{Fig3}a (black
  plus signs and error bars) is much better described by the best-fit kappa
  distribution (blue dashed line) than by the best-fit Maxwellian distribution
  (red dash-dotted line). The overall better description that the kappa
  distribution fits yield for observations throughout the entire 90-s interval
  is indicated in Figure~\ref{FigSum}d, which shows values of the reduced
  chi-squared statistic
  \begin{linenomath*}
  \begin{equation} \label{eqChi2Spec} \chi_{\mathrm{red}}^2 = \sum\limits_i^N
    \frac{1}{F} \frac{ \left ( Y_i(x) - y_i(x) \right )^2 }{w_i^2}
  \end{equation}
  \end{linenomath*}
  for 2-D fits using either a kappa distribution (blue triangles) or a
  Maxwellian distribution (red squares). In this expression $i$ indexes each
  pitch-angle and energy bin used in the fitting procedure, $Y_i$ is the
  observed differential energy flux, $y_i$ is the differential energy flux of
  the original best-fit 2-D model distribution, $w_i$ is the uncertainty due to
  counting statistics, and $F$ is the degrees of freedom, or the total number of
  pitch angle--energy bins $N$ minus the number of free model parameters. The
  difference is most pronounced after approximately 01:05:00~UT, corresponding
  to the interval during which $\kappa_{\textrm{min}} < \kappa \lesssim$~2 in
  Figure~\ref{FigSum}c.

  Figure~\ref{FigSum}e shows that over approximately the first half of the 90-s
  interval, $T =$~30--200~ev for both Maxwellian and kappa distribution
  fits. Comparison with statistical plasma sheet temperatures reported by
  \citet{Kletzing2003} (their Figure 4) indicates that these temperatures are
  within typical ranges.
  During the latter half of the interval the most probable kappa temperatures
  tend to be much higher ($\sim$100--2000~eV) than during the first half, while
  corresponding Maxwellian temperatures remain low (20--100~eV). The kappa
  ``core temperatures'' \citep{Nicholls2012} $T_c = T (1-\frac{3}{2 \kappa})$
  (not shown) during the latter half are nonetheless within several eV of
  Maxwellian temperatures.

  Figure~\ref{FigSum}f shows that relative to the calculated densities, the
  best-fit densities for Maxwellian and kappa distributions are generally within
  factors of two to five. For calculated densities the corresponding
  uncertainties (vertical black bars) are obtained using analytic expressions for
  moment uncertainties related to counting statistics for an arbitrary
  distribution function \citep{Gershman2015}. These uncertainties are generally
  less than 10\% of the calculated density.

  \subsection{Inference of magnetospheric source parameters}\label{sAfterData}

  We now demonstrate how the observed electron distributions may indicate the
  properties of the magnetospheric source region. This is performed via
  comparison of the predictions of moment-voltage relationships (\ref{eqKnight})
  and (\ref{eqDors})--(\ref{eqDensVK}) with experimental moment-voltage
  relationships derived from model-independent fluid moments of the electron
  distributions observed during the delineated interval between dashed lines in
  Figure~\ref{FigSum}, 01:04:31--01:04:41~UT. We have selected this interval
  because it is associated with the largest variation in the inferred potential
  during the entire 90-s period.

  The J-V and J$_E$-V relationships are formed by first determining the
  potential drop $\Delta \Phi$ (solid white line, Figure~\ref{FigSum}b) at each
  time, which is taken to be the peak electron energy $E_p$. (There is no
  potential drop below FAST during this interval, which would otherwise be
  indicated by the presence of upgoing ion beams; see, e.g.,
  \citealp{Elphic1998}; \citealp{Hatch2018}.) We define the peak energy $E_p$ as
  the energy of the EESA channel above which the observed differential flux
  spectrum exhibits exponential or power-law decay
  \citep{Kaeppler2014a,Ogasawara2017}, within the earthward loss cone. We then
  calculate the parallel electron current density $j_{\parallel,i}$ and energy
  flux $j_{E\parallel,i}$ of the observed electron distribution, using
  measurements from the peak energy $E_p$ up to 5~keV and the range of angles
  within the earthward loss cone. The upper bound of the energy integration
  range is limited to 5~keV because statistics of particles above this energy
  are poor and contribute almost exclusively to the uncertainty of these
  moments. These two moments are mapped to the ionosphere at 100~km using IGRF
  11 (denoted by the subscript $i$).

  The n-V relationship is also formed from the inferred potential drop
  $\Delta \Phi$ and from the calculated number density $n$, but unlike the
  fluxes $j_{\parallel,i}$ and $j_{E\parallel,i}$, $n$ is not a flux and is not
  straightforward to map to the ionosphere. We therefore must form a ``local''
  (i.e., unmapped) n-V relationship and obtain $n$ via integration over the same
  range of energies that are used to calculate $j_{\parallel,i}$ and
  $j_{E\parallel,i}$ (from $E_p$ up to 5~keV), but over a modified pitch angle
  range, which in a local treatment should be the full 180$^\circ$ range of
  earthward pitch angles. However, inspection of the delineated interval in
  Figure~\ref{FigSum}a indicates the presence of a prominent trapped population
  at pitch angles $\vert \theta \vert \gtrsim$~40$^\circ$ (e.g., at 01:04:37~UT)
  that should not be included in the calculation of $n$. We therefore integrate
  over a 60$^\circ$ range of angles that is centered on the earthward loss cone
  and multiply both the calculated densities and their uncertainties by the
  solid-angle ratio $1/(1-\cos 30^\circ) \approx 7.46$ to compensate for the
  exclusion of observations over pitch angles $\vert \theta \vert
  >$~30$^\circ$. This multiplication assumes the primary electron distribution
  is isotropic outside the loss cone.


  \begin{figure}
    \centering
    \noindent\includegraphics[width=0.90\textwidth]{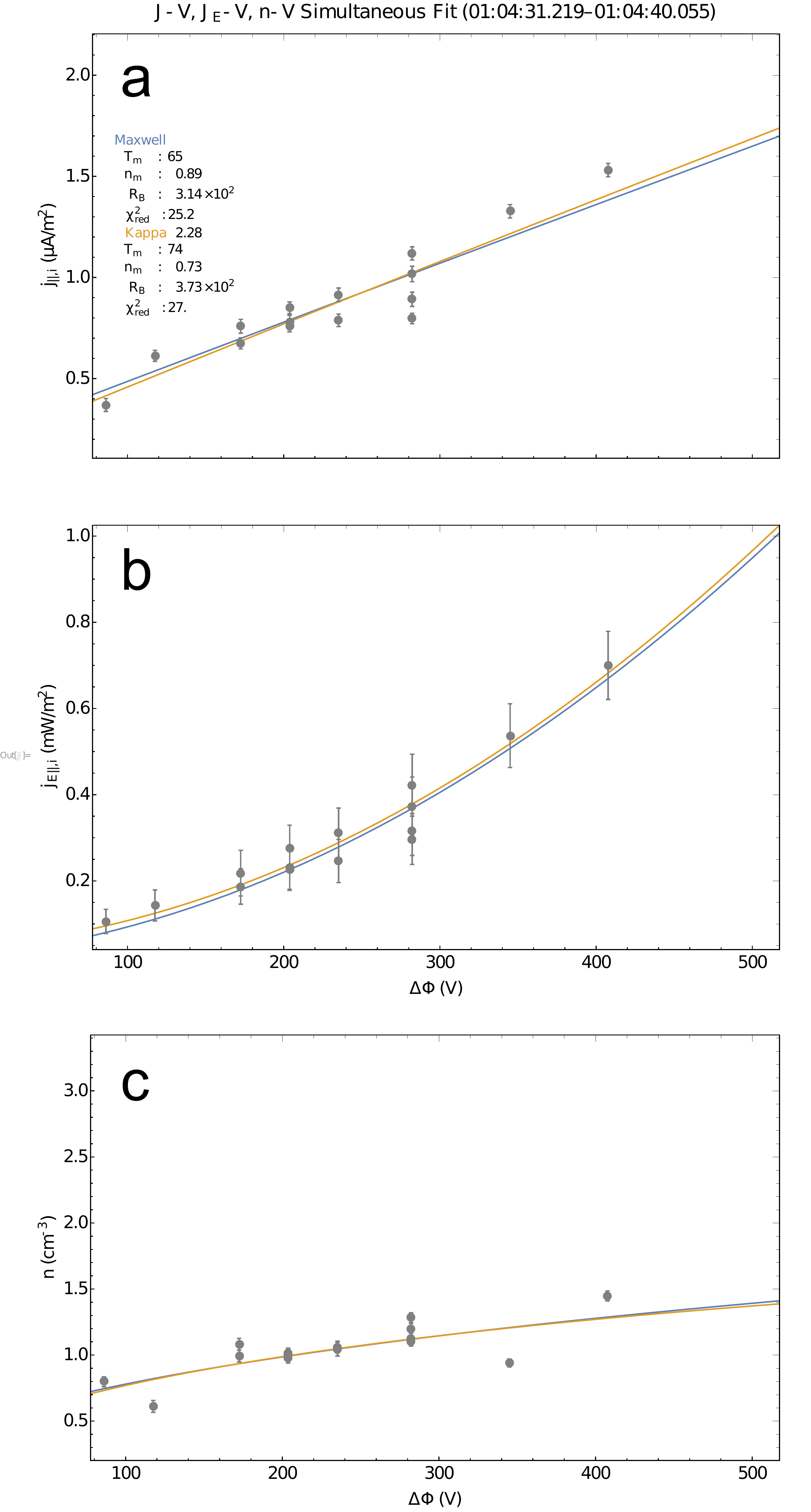}
    \caption{J-V, J$_E$-V, and n-V relationships inferred from electron
      observations during the interval 01:04:28--01:04:41~UT in
      Figure~\ref{FigSum}, together with best-fit Maxwellian (solid blue lines)
      and kappa (solid orange lines) moment-voltage relationships obtained by
      simultaneously fitting all three experimentally inferred moment-voltage
      relationships. (a) J-V relationship. (b) J$_E$-V relationship. (c) n-V
      relationship. Given values of the inferred potential drop $\Delta \Phi$
      occur multiple times within the sample interval, causing the experimental
      data to be multi-valued. Calculated current densities and energy fluxes
      (bullets) as well as their uncertainties (1$\sigma$) are mapped to the
      ionosphere at 100~km as described in the text. Calculated number densities
      are not mapped to the ionosphere. Moment uncertainties are obtained as
      analytic moments of observed electron distributions, as described in
      Appendix \ref{AppB}. The $\chi_{\mathrm{red}}^2$ values indicated in
      Figure 4a are the sum of the $\chi_{\mathrm{red}}^2$ value corresponding
      to each moment-voltage relationship.}
    \label{Fig4}
  \end{figure}


  The experimental J-V, J$_E$-V, and n-V relationships are shown in
  Figures~\ref{Fig4}a--c. Also shown are the results of simultaneously fitting
  all three of these relationships with the corresponding Maxwellian and kappa
  moment-voltage relationships (\ref{eqKnight})--(\ref{eqDensVK}) using
  NonlinearModelFit in the Mathematica\textsuperscript{\textregistered} (v11.3)
  programming language. The best-fit Maxwellian (blue lines) and kappa (orange
  lines) fits correspond to $\chi^2_{\mathrm{red}} =$~25.2 and
  $\chi^2_{\mathrm{red}} =$~27, respectively, where these two
  $\chi^2_{\mathrm{red}}$ values are the sum of the three
  $\chi^2_{\mathrm{red}}$ corresponding to each type of moment-voltage
  relation.

  We obtain these fits by drawing from random variables
  $N\sim U($0.05~cm$^{-3}$,1.5~cm$^{-3})$ and $R\sim U(5,10^4)$ to initialize
  $n_m$ and $R_B$ in each moment-voltage relationship. For the kappa
  moment-voltage relationships we must also draw from a random variable to
  initialize the $\kappa$ parameter. To accomplish this we randomly choose a
  degree of correlated motion $W \sim U(0.05,0.85)$ (see Equation~\ref{eqCorr}),
  from which we obtain an initial kappa value $K\sim \kappa_{\textrm{min}} /
  W$. The lower and upper bounds of the uniformly random degree of correlation
  $W$, $\rho = 0.05$ and $\rho = 0.85$, respectivly correspond to $\kappa = 30$
  and $\kappa = 1.76$.

  For both types of fits the parameter $T_m$ is held fixed. For the fits
  involving the Maxwellian moment-voltage relationships the value of $T_m$ is
  set equal to the median (65~eV) of the best-fit Maxwellian distribution
  temperatures (red boxes in Figure~\ref{FigSum}e) during the marked
  interval. For the fits involving the kappa moment-voltage relationships the
  value of $T_m$ is set equal to the median (74~eV) of the best-fit kappa
  distribution temperatures (blue triangles in
  Figure~\ref{FigSum}e). Additionally, because the experimental $n$ values in
  Figure~\ref{Fig4}c are not mapped to the ionosphere, the $R_B$ parameter in
  the n-V relationships (\ref{eqDensVM})--(\ref{eqDensVK}) must be reduced by a
  factor $R_{B,\textrm{FAST}} = \frac{B_{i}}{B_{\textrm{FAST}}} \approx 4$ when
  fitting the experimental n-V relationship.

  Similar to the process described at the beginning of this section for Monte
  Carlo simulation of 2-D distribution fits, to determine the range of
  parameters that may describe the observed moment-voltage relationships we
  perform fits to $N =$~2,000 Monte Carlo simulated moment-voltage relationships
  for each type of J-V, J$_E$-V, and n-V relationship, either Maxwellian or
  kappa. For each iteration, we add to each of the inferred potential drop
  values a uniform random number $X \sim \mathcal{N}(0,\Delta E_{p})$, where the
  value in the second argument is the uncertainty of the electron peak energy,
  which arises from the EESA energy channel spacing. We insert these synthetic
  potential drop values into the best-fit J-V, J$_E$-V,and n-V relationships and
  add to each of these theoretical moment predictions a normal random number
  $Z \sim \mathcal{N}(0,1)$ multiplied by the uncertainty of the corresponding
  current density, energy flux, or number density measurements. We then draw
  from the random variables $N$, $R$, and $K$, which are as described above, to
  initialize $n_m$, $R_B$, and $\kappa$, respectively. We then perform the fit.


  \begin{figure}
    \centering
    \noindent\includegraphics[width=1.0\textwidth]{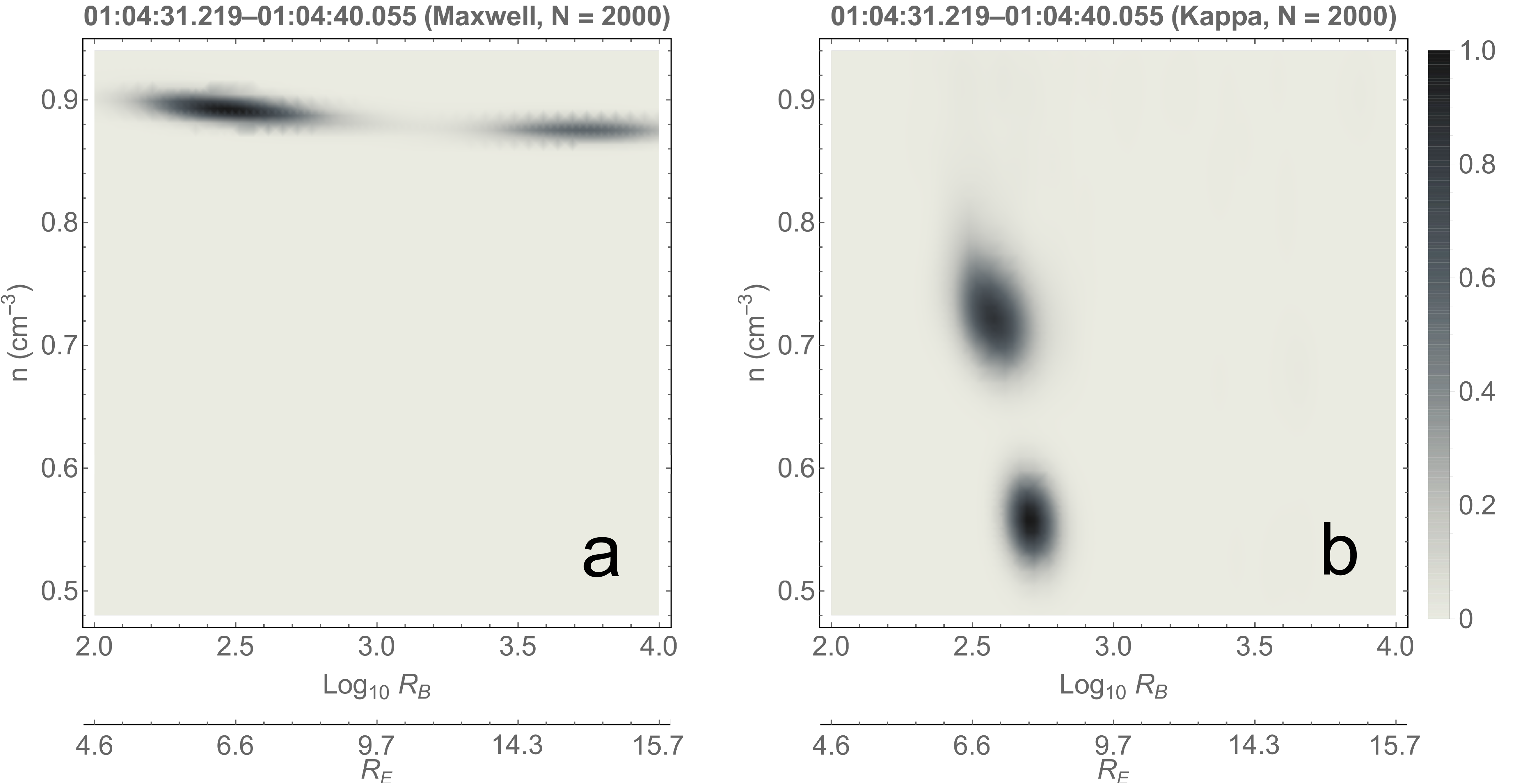}
    \caption{(a) Joint distribution of density $n_m$ and mirror ratio $R_B$ for
      Maxwellian fits of $N =$~2,000 Monte Carlo simulated J-V, J$_E$-V, and n-V
      relationships (\ref{eqKnight}), (\ref{eqJEM}), and (\ref{eqDensVM}). (b)
      Same as Figure~\ref{Fig5}a, except that fits are performed using the kappa
      J-V, J$_E$-V, n-V relationships (\ref{eqDors}), (\ref{eqJEK}), and
      (\ref{eqDensVK}). In both panels the $R_B$ axis is logarithmic and the
      secondary axis shows the approximate source height $h$ in Earth radii. The
      gray scale indicates the distribution height in units such that the peak
      value of each distribution is 1.0.}
    \label{Fig5}
  \end{figure}


  The resulting joint distributions of $n_m$ and $R_B$ are shown in
  Figures~\ref{Fig5}a and \ref{Fig5}b for the Maxwellian and kappa
  moment-voltage relationships, respectively. The Maxwellian moment-voltage
  relationships predict two different solution regimes: the first corresponds to
  $n_m =$~0.88--0.90~cm$^{-3}$ and $R_B =$~200--500; the second corresponds to
  $n_m =$~0.87--0.88~cm$^{-3}$ and $R_B =$~3,200--7,500. The secondary axis
  indicates the approximate source altitudes $h =$5.7--7.7~$R_E$ and
  $h =$14--15.5~$R_E$, respectively, where $R_E$ indicates the radius of
  Earth. The kappa moment-voltage relationships also show two different solution
  regimes: the first corresponds to $n_m =$~0.70--0.79~cm$^{-3}$,
  $R_B =$~300---510 ($h =$6.4--7.7~$R_E$), and $\kappa =$~2.2--2.8; the second
  corresponds to $n_m =$~0.55--0.56~cm$^{-3}$, $R_B =$~300--500
  ($h =$6.4--7.6~$R_E$), and $\kappa \leq$~1.8.

  The $\chi^2_{\mathrm{red}}$ value for the kappa fits in Figure~\ref{Fig4} is
  7\% greater than the $\chi^2_{\mathrm{red}}$ value for the Maxwellian fits. It
  therefore seems impossible to determine the correct solution regime solely on
  the basis of information in Figures~\ref{Fig4} and \ref{Fig5}. However, only
  the first kappa solution regime is consistent with the range $\kappa =$~2--9
  that arises from direct 2-D distribution fits during the 10-s delineated
  period in Figure~\ref{FigSum}c. 

  We have also performed $N = 6,000$ Monte Carlo simulations using only the
  inferred J-V relationship (Figure~\ref{Fig4}a) and either the the Maxwellian
  J-V relation (\ref{eqKnight}) or the kappa J-V relation (\ref{eqDors}). From
  the Maxwellian J-V relation we obtain solutions corresponding to
  $n_m =$~0.88--0.90~cm$^{-3}$ and $R_B \geq$~680 ($h \geq$8.5~$R_E$). From the
  kappa J-V relation (\ref{eqDors}) we obtain solutions corresponding to
  $n_m =$~0.68--0.84~cm$^{-3}$ and $R_B \geq$~2,100 ($h \geq$12.8~$R_E$) for
  $\kappa =$~2.2--2.8. Thus for this case study and an assumed Maxwellian or
  kappa source population, the source altitude lower bound is relatively greater
  when only the J-V relationship is used.
 
  \section{Orbit 4682}


  \begin{figure}
    \centering
    \noindent\includegraphics[width=1.0\textwidth]{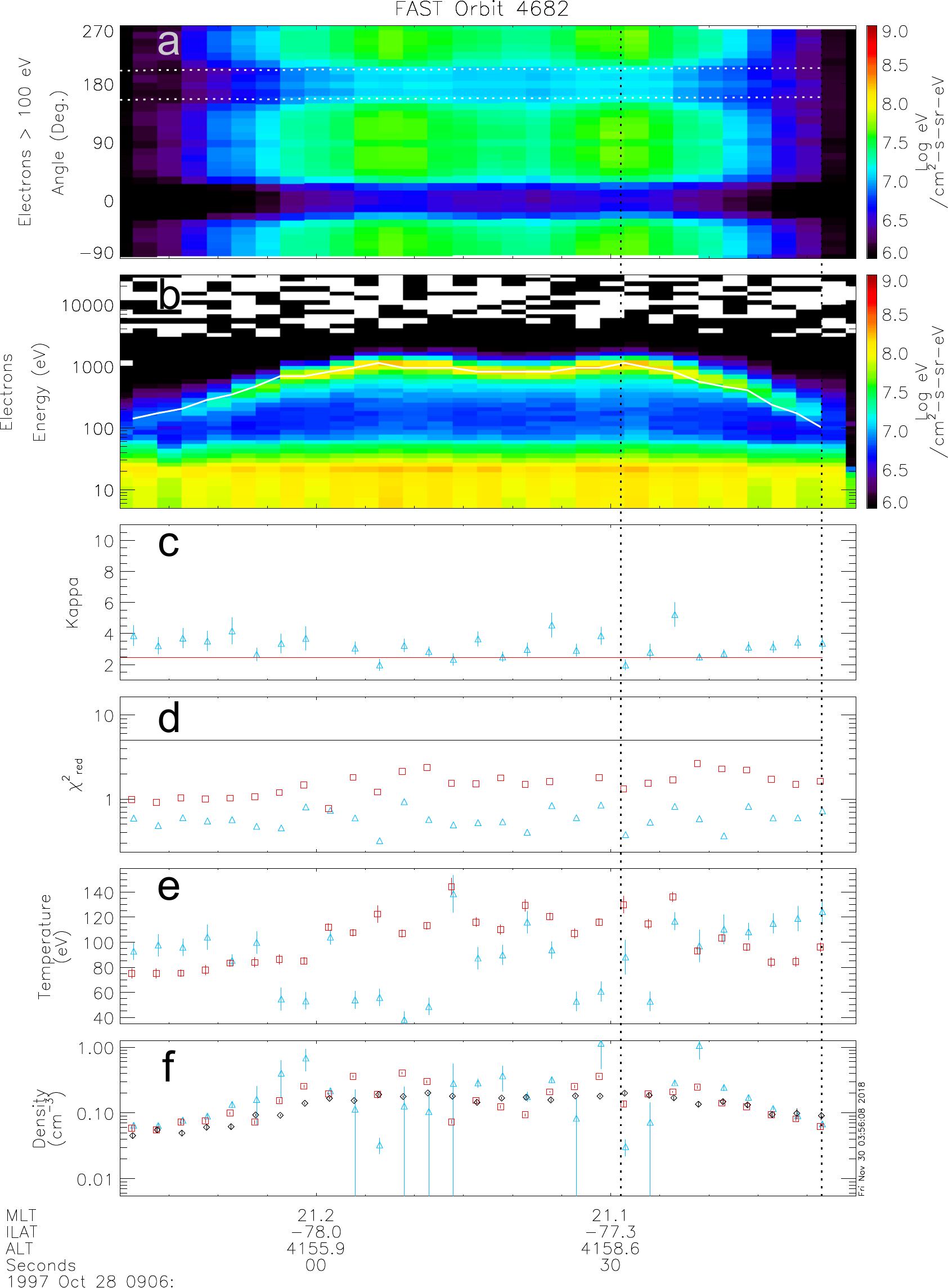}
    \caption{EESA observations of inverted V precipitation on Oct 28, 1997 and
      corresponding 2-D fit parameters, in the same format as
      Figure~\ref{FigSum}. (a) $>$100~eV electron pitch-angle distribution. The
      earthward portion of the loss cone comprises the range of pitch angles
      between dotted horizontal lines at $\sim$150$^\circ$ and
      $\sim$210$^\circ$.  (b) Average electron energy spectrum within the
      earthward loss cone. (c) $\kappa$ fit parameter for the best-fit kappa
      distribution. (d) Reduced chi-squared statistic $\chi^2_{\mathrm{red}}$
      for each fit type. (e) Best-fit temperatures. (f) Calculated and best-fit
      densities. Calculated densities are also obtained as model-independent
      moments via integration over from the energy of the channel immediately
      below $E_p$ up to 5~keV, and over all pitch angles
      $\vert \theta \vert >$~150$^\circ$. Uncertainties of calculated densities
      and best-fit density and temperature parameters are obtained as described
      in the Figure~\ref{FigSum} caption. EESA observations during this interval
      have a sample period $T = 2.5$~s.}
    \label{FigSum2}
  \end{figure}


  \subsection{Data Presentation} 

  During a 75-s interval on Oct 28, 1997, the FAST satellite observed inverted V
  electron precipitation (Figures~\ref{FigSum2}a and \ref{FigSum2}b) at
  $\sim$21~MLT and -78$^\circ$~invariant latitude in the Southern Hemisphere
  during moderately low geomagnetic activity ($K_p
  =$~2). 
  The pitch-angle spectrogram in Figure \ref{FigSum2}a shows that precipitation
  within the earthward loss cone (range of pitch angles between horizontal
  dotted white lines in Figure~\ref{FigSum2}a) is weak
  ($dJ_E/dE\lesssim 5\times 10^7$eV/cm$^2$-s-sr-eV), while trapped electrons
  over 30$^\circ\lesssim \vert \theta \vert \lesssim$~150$^\circ$ are relatively
  more intense. Over the entire 75-s interval as well as over the $\sim$20-s
  period between dashed lines, 09:06:31--09:06:51.5~UT, Figure~\ref{FigSum2}b
  and c respectively show $E_p =$~100--1200~eV and $\kappa
  \approx$~2--5. Figure~\ref{FigSum2}d shows that best-fit Maxwellian
  $\chi^2_{\mathrm{red}}$ values are generally twice or more those of best-fit
  kappa $\chi^2_{\mathrm{red}}$ values.

  Best-fit temperatures shown in Figure~\ref{FigSum2}e indicate that over the
  entire interval $T =$~75--130~ev for Maxwellian distribution fits, while
  $T =$~35--145~ev for kappa distribution fits. As with temperatures in
  Figure~\ref{FigSum}e, these ranges of temperatures are within the typical
  range for plasma sheet electrons.

  Densities calculated directly from observed electron distributions in
  Figure~\ref{FigSum2}f (black diamonds) are within the range
  0.01--0.5~cm$^{-3}$ that is typically observed in the distant plasma sheet
  \citep{Kletzing2003,Paschmann2003}. Most probable Maxwellian and kappa fit
  densities in Figure~\ref{FigSum2}f tend to be with factors of 2 of the
  calculated densities. Similar to the density moments and uncertainties
  calculated in the previous section, we calculate the density over all energies
  from $E_p$ up to 5~keV and over all earthward pitch angles in the Southern
  Hemisphere, $\vert \theta \vert >$~150$^\circ$. We multiply calculated
  densities and density uncertainties by the solid-angle ratio
  $1/(1-\cos 30^\circ) \approx 7.46$ to compensate for the exclusion of primary
  electrons over the range of downgoing pitch angles dominated by trapped
  electrons ($90^\circ< \theta <$~150$^\circ$ and
  $-$150$^\circ< \theta <-90^\circ$).


  \begin{figure}
    \centering
    \noindent\includegraphics[width=1.0\textwidth]{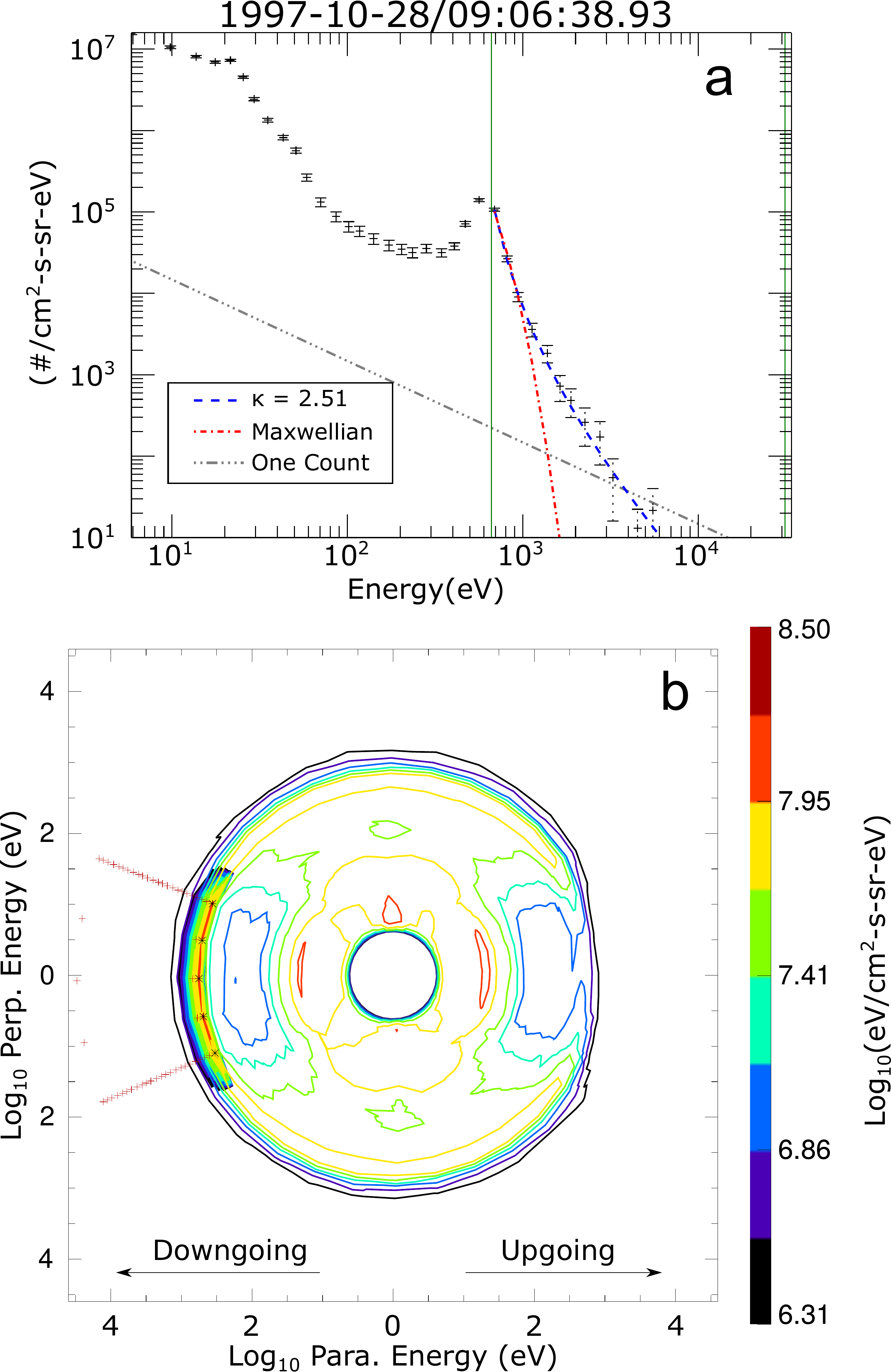}
    \caption{Electron spectra observed at 09:06:38.931--09:06:41.437~UT. The
      layout is the same as that of Figure~\ref{Fig3}. (a) 1-D differential
      number flux spectrum (black crosses) obtained by averaging all
      differential number flux spectra over all pitch angles within the
      earthward loss cone, with best-fit Maxwellian and kappa distributions
      overlaid (red dash-dotted line and blue dashed line, respectively). (b)
      Best-fit 2-D kappa distribution (solid contours) with the observed 2-D
      differential energy flux spectrum overlaid (contour lines). For each pitch
      angle black asterisks indicate the peak energy $E_p = 560$~eV, and red
      plus signs outline the range of energies and pitch angles used to perform
      the 2-D fit.}
    \label{Fig7}
  \end{figure}



  Figure~\ref{Fig7} shows an example of the electron distributions observed
  during the delineated period (09:06:31--09:06:51.5~UT), in the same layout as
  Figure~\ref{Fig3}. As in Figure~\ref{Fig3}, the best-fit kappa distribution
  (blue dashed line) successfully describes the suprathermal tail, and is a
  better fit than the Maxwellian distribution (red dash-dotted line) as
  reflected in the $\chi^2_{\mathrm{red}}$ values, respectively 0.56 and 2.57
  (also Figure~\ref{FigSum2}d).

  \subsection{Inference of magnetospheric source parameters}\label{sAfterData2}


  \begin{figure}
    \centering
    \noindent\includegraphics[width=0.90\textwidth]{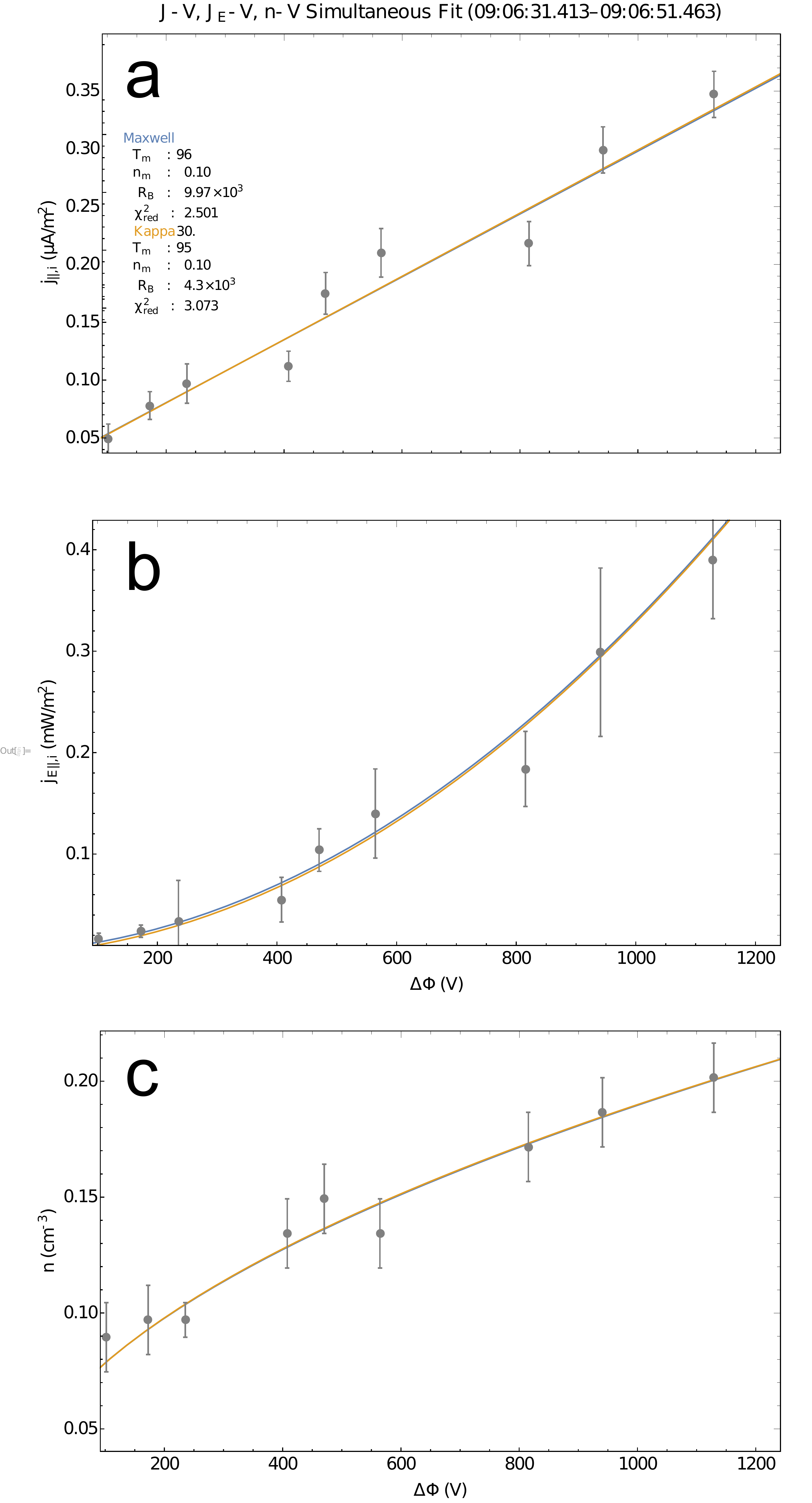}
    \caption{J-V, J$_E$-V, and n-V relationships inferred from electron
      observations during the interval 09:06:31--09:06:51.5~UT in
      Figure~\ref{FigSum2}, together with best-fit Maxwellian (solid blue lines)
      and kappa (solid orange lines) moment-voltage relationships obtained by
      simultaneously fitting all three experimentally inferred moment-voltage
      relationships. The format is the same as that of Figure~\ref{Fig4}. (a)
      J-V relationship. (b) J$_E$-V relationship. (c) n-V
      relationship. Calculated current densities and energy fluxes (black plus
      signs) as well as their uncertainties (1$\sigma$) are mapped to the
      ionosphere at 100~km as described in section~\ref{sAfterData}. Calculated
      number densities are not mapped to the ionosphere. Moment uncertainties
      are obtained as analytic moments of observed electron distributions
      (Appendix \ref{AppB}). The $\chi_{\mathrm{red}}^2$ values indicated in
      Figure 8a are the sum of the $\chi_{\mathrm{red}}^2$ values corresponding
      to each moment-voltage relationship.}
    \label{Fig8}
  \end{figure}


  Using the Monte Carlo simulation process described in
  section~\ref{sAfterData}, we now determine the range of parameters that may
  describe the observed moment-voltage relationships during the 20-s period
  shown between dashed lines in Figure~\ref{FigSum2} assuming each type of
  magnetospheric source population, either Maxwellian or kappa. We select this
  period because the inferred potential drop (solid white line in
  Figure~\ref{FigSum2}b) decreases by roughly an order of magnitude, from
  $\sim$1150~eV to $\sim$150~eV.

  The experimental J-V, J$_E$-V, and n-V relationships are shown in
  Figures~\ref{Fig8}a--c. Also shown are the results of simultaneously fitting
  all three of these relationships with the corresponding Maxwellian and kappa
  moment-voltage relationships (\ref{eqKnight}),
  (\ref{eqDors})--(\ref{eqDensVK}). Best-fit Maxwellian (blue lines) and kappa
  (orange lines) fits respectively correspond to $\chi^2_{\mathrm{red}} =$~2.5
  and $\chi^2_{\mathrm{red}} =$~3.1. As in section~\ref{sAfterData}, we obtain
  these fits by drawing from random variables
  $N\sim U($0.01~cm$^{-3}$,0.5~cm$^{-3}$), $R\sim U(5,10^4),$ and
  $K\sim \kappa_{\textrm{min}} / W$ to initialize $n_m$, $R_B$, and $\kappa$ in
  each moment-voltage relationship. (The random variable $K$ is only used in the
  kappa moment-voltage relationships.) For both types of fits the parameter
  $T_m$ is held fixed. For the fits involving the Maxwellian moment-voltage
  relationships the value of $T_m$ is set equal to the median $T =$~96~eV of the
  best-fit Maxwellian distribution temperatures (red boxes in
  Figure~\ref{FigSum2}e) during the marked interval. For the fits involving the
  kappa moment-voltage relationships the value of $T_m$ is set equal to the
  median $T =$~95~eV of the best-fit kappa distribution temperatures (blue
  triangles in Figure~\ref{FigSum2}e).


  \begin{figure}
    \centering
    \noindent\includegraphics[width=0.90\textwidth]{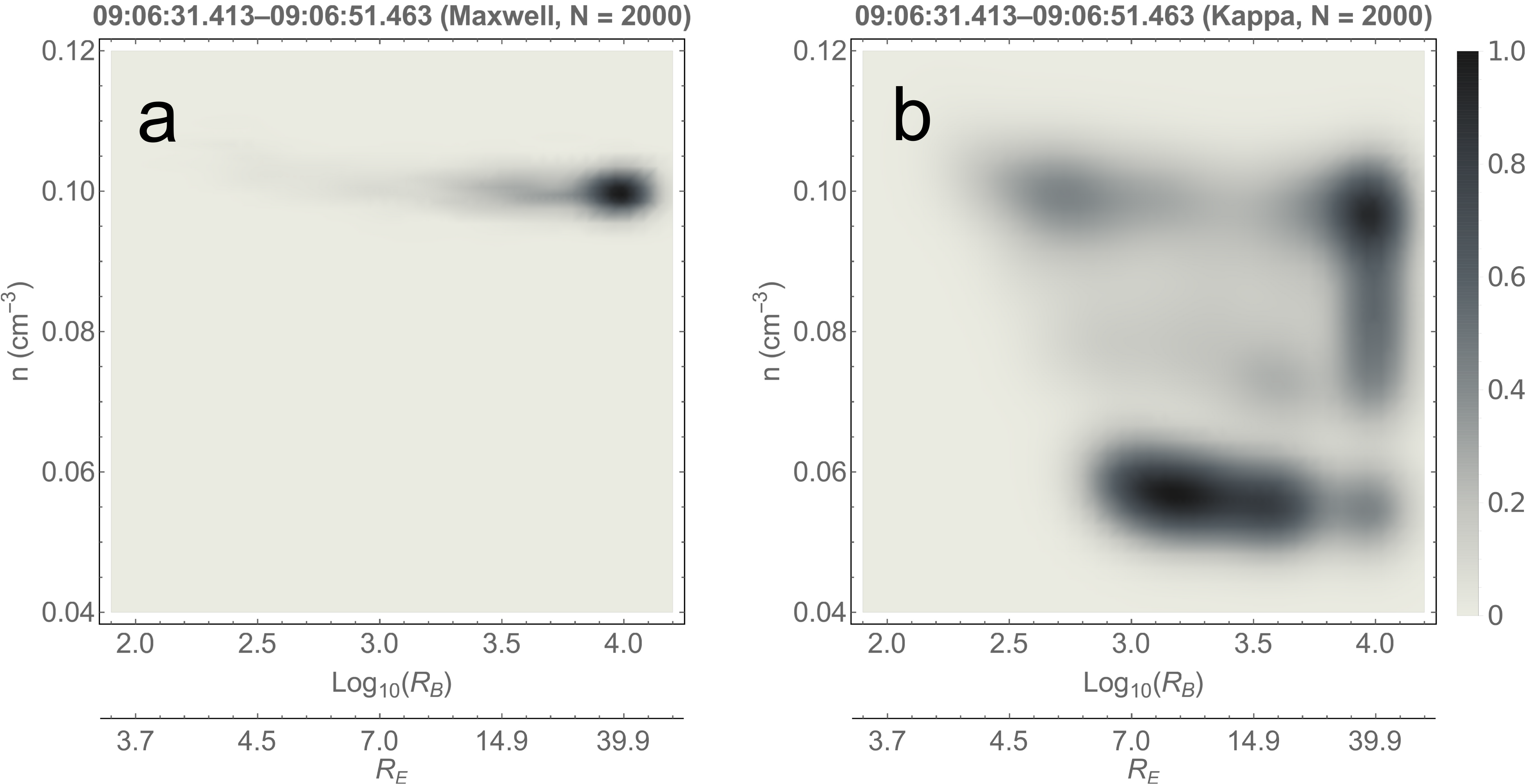}
    \caption{(a) Joint distribution of density $n_m$ and mirror ratio $R_B$ for
      Maxwellian fits of $N =$~2,000 Monte Carlo simulated J-V, J$_E$-V, and n-V
      relationships (\ref{eqKnight}), (\ref{eqJEM}), and (\ref{eqDensVM}). (b)
      Same as Figure~\ref{Fig9}a, except that fits are performed using the kappa
      J-V, J$_E$-V, n-V relationships (\ref{eqDors}), (\ref{eqJEK}), and
      (\ref{eqDensVK}). In both panels the $R_B$ axis is logarithmic and the
      secondary axis shows the approximate source height $h$. The gray scale
      indicates the distribution height in units such that the peak value of
      each distribution is 1.0.}
    \label{Fig9}
  \end{figure}

  The resulting joint distributions of $n_m$ and $R_B$ is shown in
  Figures~\ref{Fig9}a and \ref{Fig9}b for the Maxwellian and kappa
  moment-voltage relationships, respectively, in a layout identical to that of
  Figure~\ref{Fig5}. The Maxwellian moment-voltage relationships predict
  $n_m =$~0.097--0.103~cm$^{-3}$ and $R_B \geq$1,400 ($h \geq$~8.1~$R_E$). The
  kappa moment-voltage relationships show several different solution regimes,
  all of which correspond to $R_B \gtrsim$~370 ($h \gtrsim$~4.6~$R_E$) and
  $n_m =$~0.055--0.10~cm$^{-3}$: the first overlaps with the Maxwellian solution
  regime in Figure~\ref{Fig9}a, and corresponds to
  $n_m =$~0.096--0.0102~cm$^{-3}$, $R_B \geq$~370 ($h \geq 4.6$), and
  $\kappa > 10$; the second corresponds to $n_m =$~0.072--0.094~cm$^{-3}$,
  $R_B \gtrsim$~700 ($h \gtrsim 5.7$), and 2~$\leq \kappa \geq$~10; the third
  corresponds to $n_m =$~0.054--0.059~cm$^{-3}$, $R_B \gtrsim$~10$~3$
  ($h \gtrsim 6.8$), and $\kappa <$~2.

  The $\chi^2_{\mathrm{red}}$ value for the kappa fits in Figure~\ref{Fig8} is
  $\sim$24\% greater than the $\chi^2_{\mathrm{red}}$ value for the Maxwellian
  fits. As with results in the previous section, information in
  Figures~\ref{Fig8} and \ref{Fig9} seems insufficient to determine the correct
  solution regime. However, only the second kappa solution regime is consistent
  with the direct 2-D distribution fits during the 20-s delineated period in
  Figure~\ref{FigSum2}.

  As in section~\ref{sAfterData} we have performed $N = 6,000$ Monte Carlo
  simulations using only the inferred J-V relationship (Figure~\ref{Fig8}a)
  and either the the Maxwellian J-V relation or the kappa J-V relation. From the
  Maxwellian J-V relation the resulting solutions correspond to
  $n_m =$~0.097--0.106~cm$^{-3}$ and $R_B \geq$~1,900 ($h \geq$10~$R_E$). From
  the kappa J-V relation (\ref{eqDors}) we obtain solutions corresponding to
  $n_m =$~0.071--0.093~cm$^{-3}$ and $R_B \geq$~1,500 ($h \geq$8.6~$R_E$) for
  $\kappa =$~2--10. Similar to results in section~\ref{sAfterData}, the source
  altitude lower bound is relatively greater when only the J-V relationship is
  used.
 
  \section{Discussion and Summary}
  \label{sDiscussion}

  \begin{table}
    \caption{Most likely magnetospheric source parameters}
    \centering
    \begin{tabular}{l c c c c c c}
      \hline
                  & Source Type & Temperature$^{a}$  & Density      & $R_B$          & $h$      & $\kappa$ \\
                  &             &  (eV)              & (cm$^{-3}$)   &                & ($R_E$)  &         \\
      \hline
      Orbit 1607  & Maxwellian & 65                  & 0.87--0.9    & 220--6,400     & 5.8--15  &          \\
                  & Kappa      & 74                  & 0.70--0.79   & 300--510       & 6.4--7.7 & 2.2--2.8  \\
      Orbit 4682  & Maxwellian & 96                  & 0.097--0.103 & $\geq$1400     & $\geq$8.1 &         \\
                  & Kappa      & 95                  & 0.071--0.091 & $\geq$720      & $\geq$5.9 & 2--6    \\
      \hline
      \multicolumn{7}{l}{$^{a}$Fixed.}
    \end{tabular}
    \label{Tab1}
  \end{table}

  For the two case studies that we have presented we assume either Maxwellian or
  kappa source populations when fitting the observed J-V, J$_E$-V, and n-V
  relationships, which results in $\chi^2_{\mathrm{red}}$ values that differ by
  a few to several percent (see Figures~\ref{Fig4} and \ref{Fig8}). Such
  differences indicate that the moment-voltage relationships themselves are
  insufficient to determine the source region properties. We identify the most
  likely ranges of source densities and altitudes in each case study by
  requiring that these parameters correspond to the range of $\kappa$ values
  estimated from direct 2-D fits of observed electron distributions.

  Table~\ref{Tab1} summarizes the ranges of most likely source parameters for
  both case studies. As stated in previous sections, the estimated temperatures
  and densities are within or near the typical ranges expected on the basis of
  surveys of the plasma sheet. The combined range of $\kappa$ values estimated
  for each orbit, $\kappa =$~2--6, are also within the ranges indicated by
  in situ plasma sheet surveys
  \citep{Christon1989,Christon1991,Kletzing2003,Stepanova2015}.

  The estimated ranges of source altitudes for these two case studies,
  $h =$~6.4--7.7 for Orbit 1607 observations and $h \geq$~5.9~$R_E$ for Orbit
  4682 observations, are above the typically quoted range of altitudes
  $\sim$1.5--3~$R_E$ for the auroral acceleration region
  \citep{Mozer2001,Morooka2004,Marklund2011}. Results from previous studies
  \citep{Wygant2002,Li2014} indicate that such ``high-altitude acceleration''
  scenarios often involve Alfv\'{e}n wave-particle interactions, and
  \citet{Andersson2002a} have shown that the signatures of these interactions at
  high altitudes may appear monoenergetic.

  There are three primary limitations of this study. First, verification of the
  results shown in Table~\ref{Tab1} requires conjunctive observations along
  similar field lines from FAST in the acceleration region and from another
  spacecraft in the source region; unfortunately the latter are not available
  during the intervals shown in Figures~\ref{FigSum} and \ref{FigSum2}. Second,
  related to the previous point, the clear monoenergetic peaks in
  Figures~\ref{FigSum}b and \ref{FigSum2}b suggest that the potential structures
  above FAST are stationary relative to the transit time of plasma sheet
  electrons. We nevertheless cannot directly verify our assumption that
  quasistatic magnetospheric processes and monotonic potential structures are
  the cause of the electron precipitation shown in Figures~\ref{FigSum}a--b and
  Figures~\ref{FigSum2}a--b. Third, the moment-voltage relationships
  (\ref{eqKnight}) and (\ref{eqDors})--(\ref{eqDensVK}) assume that the
  magnetospheric source population is isotropic.

  Studies performed by \citet{Hull2010} and \citet{Marklund2011} have shown that
  potential structures are generally neither quasistatic nor monotonic, and
  \citet{Hatch2018} present statistics suggesting that electron distributions
  may be modified in the vicinity of the AAR. These studies indicate that our
  assumptions of stationarity, non-variability of source parameters along the
  mapped satellite track, and adiabatic transport from the source region to the
  ionosphere are not always true, and some evidence of violation of our
  assumptions appears in, for example, the experimental J-V relation (top panel)
  in Figure~\ref{Fig4}: For some data points neither the Maxwellian nor the
  kappa J-V relation is within 2--3$\sigma$. Such differences could suggest that
  the errors associated with our assumptions are larger than that associated
  with moment uncertainty and counting statistics.

  Concerning the third limitation, magnetospheric source populations are not 
  necessarily isotropic, and previous studies \citep{Marghitu2006,Forsyth2012}
  have shown how the observed degree of anisotropy of electron precipitation may
  in fact be used to estimate the source altitude. While outside the scope of
  the present study, relaxing the assumption of isotropy and adapting the source
  altitude estimation techniques presented by these previous studies are natural
  future extensions of the techniques we have developed for the two case studies
  presented above.

  Regardless of the particular values or ranges of parameters that we have
  identified, these case studies nevertheless demonstrate how the non-Maxwellian
  nature of an electron source population may be embedded in the observed
  moment-voltage relationships, requiring modification of both the inferred
  source density and mirror ratio. From this standpoint the degree to which a
  source population departs from thermal equilibrium, as indicated by the
  $\kappa$ parameter in this study, is as fundamental a plasma property as
  density or temperature.

  A relatively small number of studies, such as those of
  \citet{Shiokawa1990,Lu1991,Morooka2004,Dombeck2013}, has compared various
  forms of the Knight relation (\ref{eqKnight}) to observations. To our
  knowledge, however, no study besides the present has used the moment-voltage
  relationships (\ref{eqKnight}) and (\ref{eqDors})--(\ref{eqDensVK}) that are
  predicted by Liouville's theorem, or any subset thereof, to infer the
  properties of the magnetospheric source region on the basis of observations at
  lower altitudes.

  In summary, in this study we have (i) derived the two previously unpublished
  n-V relationships (\ref{eqDensVM}) and (\ref{eqDensVK}); (ii) inferred the
  properties of magnetospheric source populations in two case studies based on
  simultaneous fitting of the three experimental moment-voltage relationships
  with corresponding theoretical moment-voltage relationships (\ref{eqKnight}),
  (\ref{eqDors})--(\ref{eqDensVK}), moment uncertainties, and direct 2-D fits of
  observed precipitating electron distributions; (iii) demonstrated that
  knowledge of the degree to which monoenergetic precipitation departs from
  Maxwellian form, which we parameterize via the $\kappa$ index, is required to
  determine the most likely set of magnetospheric source parameters.

  \appendix

  \section{Theory of collisionless transport through a field-aligned monotonic
    potential structure}


  Here we review the theory that yields the J-V, J$_E$-V, and n-V relationships
  (\ref{eqKnight}) and (\ref{eqDors})--(\ref{eqDensVK}). The development is
  intended to be brief since several more elaborate developments have been given
  elsewhere (e.g., references in the Introduction).

  Assuming a gyrotropic, collisionless magnetospheric source population, an
  electron distribution function $f(v_{\parallel},v_{\perp})$ can be written in
  terms of total energy and the first adiabatic invariant:
  \begin{linenomath*}
    \begin{subequations}\label{eqInvariants}
      \begin{align}
        \begin{split}
          E &= \frac{m_e}{2} ( v_\parallel^2 + v_\perp^2 ) + \Pi(B);
        \end{split}\label{eqEnergyCons}\\
        \begin{split}
          \mu &= \frac{m_e v_\perp^2}{2 B(s)}.
        \end{split}\label{eqMu}
      \end{align}
    \end{subequations}
  \end{linenomath*}
  In these expressions $v_\parallel$ and $v_\perp$ are parallel and
  perpendicular velocity, $\Pi(B)$ is the distribution of potential energy along
  the field line, $B(s)$ is magnetic field strength, and $s$ is a one-to-one
  function of $B$ that measures the distance along a magnetic field line from
  the magnetospheric source region toward the ionosphere. We denote the magnetic
  field strength at the source $B_m \equiv B(s_0)$ and assume $\Pi(B_m) =
  0$. Thus the initial total energy is
  $E_0 = \frac{m_e}{2} ( v_{\parallel,0}^2 + v_{\perp,0}^2 )$. From Equations
  (\ref{eqInvariants}) we then have
  $v_{\parallel,0}^2 = v_{\parallel}^2 + v_{\perp}^2
  \left(1-\frac{B_m}{B(s)}\right) + \frac{2}{m_e}\Pi(B)$, with $v_{\perp,0}^2$
  eliminated via Equation (\ref{eqMu}).


  In principle derivation of a moment-voltage relationship involves simple
  application of Equations (\ref{eqInvariants}) in Liouville's theorem,
  $f(v_\parallel,v_\perp) = f(v_{\parallel,0},v_{\perp,0})$, followed by
  calculation of the relevant moment. In practice the complexity of these
  calculations is related to the shape of $\Pi(B)$, since multiple regions of
  phase space may be inaccessible, or ``forbidden,'' at lower altitudes. (For
  example, particles with parallel velocities that are too low to overcome a
  retarding potential structure will be reflected.) Care must be taken to
  exclude such forbidden regions from moment calculations
  \citep{Liemohn1998,Bostrom2003a,Bostrom2004,Pierrard2007a}. We are interested
  in the simplest non-trivial case, namely that for which $\Pi(B)$ obeys the
  conditions $\frac{d \Pi}{dB} < 0$ and $\frac{d^2 \Pi}{dB^2} > 0$, with the
  derivatives defined everywhere along the magnetic field line.  For this case
  each moment-voltage relationship is independent of the shape of $\Pi(B)$ and
  can be written as a function of the total potential difference $\Delta \Phi$
  (e.g., the J-V relationships (\ref{eqKnight}) and (\ref{eqDors}); see
  \citealp{Liemohn1998}).

  The allowed region of phase space is $v_{\parallel} \geq 0$. Via the two
  invariants in Equation (\ref{eqInvariants}) the lower bound of this inequality
  may be written in terms of total kinetic energy $W$, initial parallel kinetic
  energy $W_{\parallel,0}$, and pitch angle
  $\theta \equiv \textrm{tan}^{-1} ( v_\perp / v_\parallel)$ as
  \begin{linenomath*}
    \begin{equation*}
      W_{\parallel,0} = W \left(1-\sin^2\left(\theta\right) \frac{B_m}{B(s)}\right) - e \Delta \Phi = 0.
    \end{equation*}
  \end{linenomath*}
  The region of phase space over which to integrate is then defined by the
  inequalities
  \begin{linenomath*}  
  \begin{subequations}\label{eqVelSpace}
      \begin{align}
        \begin{split}\label{eqWCondition}
          W &\geq e \Delta \Phi / \left(1-\sin^2\left(\theta\right)\frac{B_m}{B(s)}\right); \\
        \end{split}\\
        \begin{split}
          \theta \in \begin{cases} (-90^\circ,90^\circ) & \text{Northern Hemisphere};\\
            (90^\circ,270^\circ) & \text{Southern Hemisphere}.\\
          \end{cases}
        \end{split}
      \end{align}
    \end{subequations}
  \end{linenomath*}



  Assuming gyrotropy the zeroth moment of $f(v_{\parallel},v_{\perp})$ is
  $n = 2 \pi \iint v_{\perp} \, f (v_{\parallel},v_{\perp}) dv_{\perp}
  dv_{\parallel}$.
  The Maxwellian and kappa n-V relations (\ref{eqDensVM}) and (\ref{eqDensVK})
  result from evaluation of this integral over the boundaries (\ref{eqVelSpace})
  using either an isotropic Maxwellian or isotropic kappa distribution function,
  respectively, and assuming a total potential drop $\Delta \Phi$.



  For 1~$\lesssim \overline{\phi} \ll R_B$ the Maxwellian n-V relation
  (\ref{eqDensVM}) reduces to
  \begin{linenomath*}
  \begin{equation}\label{eqDensVMlittlePhi}
    n/n_m =  \frac{1}{2} e^{\overline{\phi}}
    \textrm{erfc}\, \overline{\phi}^{\frac{1}{2}} + \left(\overline{\phi}/\pi\right)^{\frac{1}{2}}, 
  \end{equation}
  \end{linenomath*}
  while in the limit $\overline{\phi} \gg R_B >$~1 the Maxwellian n-V relation
  reduces to
  \begin{linenomath*}
  \begin{equation}\label{eqDensVMlittleRB}
    n/n_m = \frac{1}{2} e^{\overline{\phi}}
    \textrm{erfc}\, \overline{\phi}^{\frac{1}{2}} + \frac{1}{2} \frac{R_B-1}{\left(\pi \overline{\phi}\right)^{\frac{1}{2}}}.
  \end{equation}
  \end{linenomath*}
  For fixed $R_B$ the maximum value of $n/n_m$ is given by $\overline{\phi}$
  such that
  \begin{linenomath*}
  \begin{equation}\label{EqMaxValWRTPhi}
    e^{\overline{\phi}} \textrm{erfc} \left( \sqrt{\overline{\phi}} \right) = \frac{2}{\sqrt{\pi \left(R_B-1\right)}} D \left(\sqrt{\frac{\overline{\phi}}{R_B-1}}\right).
  \end{equation}
  \end{linenomath*}

  \section{Analytic expressions for moment uncertainties}\label{AppB}

  We follow the \citet{Gershman2015} framework for calculating the moment
  uncertainties used in Monte Carlo simulations in sections \ref{sAfterData} and
  \ref{sAfterData2}. Let $W$ be a
  differentiable function of plasma moments $\langle n A_i \rangle $; the
  linearized uncertainty $\sigma_W$ may be expressed
  \begin{linenomath*}
  \begin{equation} \label{eqLinUnc} \sigma_W^2 = \sum\limits_{i}\sum\limits_{j}
    \left ( \frac{ \partial W}{\partial A_i} \right ) \left ( \frac{ \partial
      W}{\partial A_j} \right ) \sigma_{A_i,A_j}.
  \end{equation}
  \end{linenomath*}
  %
  %
  %
  %
  The squared uncertainty of $n$ is trivially $\sigma_n^2$, while squared
  uncertainties of $j_{\parallel}$ and $j_{E\parallel}$ are
  \begin{linenomath*}
    \begin{subequations}\label{eqUncerts}
    \begin{align}
      \begin{split}
        \sigma_{j_\parallel}^2 =& \, V_\parallel \sigma_{n}^{2} + n
        \sigma_{V_\parallel }^{2} + \sigma_{n,V_\parallel};
        \end{split}\label{eqjParUnc}\\
      \begin{split}
        \sigma_{j_{E\parallel}}^2 =& \, \sigma_{H,\parallel}^2 + B \left [ 2 \sigma_{H_\parallel,V_\parallel} + B \sigma_{V_\parallel}^2 \right ] \\
        &+ V_\parallel \left [ 3 \sigma_{H_\parallel, P_\parallel } + 2 \sigma_{H_\parallel, P_\perp } + B \left (3 \sigma_{V_\parallel, P_\parallel} + 2 \sigma_{ V_\parallel, P_\perp} \right ) \right ] \\
        &+ V_\parallel^2 \left [ \frac{9}{4} \sigma_{ P_\parallel }^2 +
          3 \sigma_{P_\parallel ,P_\perp} + \sigma_{P_\perp}^2 \right ];
        \end{split}\label{eqEFluxUnc}
    \end{align}
    \end{subequations}
  \end{linenomath*}
  where $V_\parallel$ is the average parallel velocity, and
  $B = \left ( \frac{3}{2} P_\parallel + P_\perp \right )$ in expression
  (\ref{eqEFluxUnc}).  Equation (\ref{eqEFluxUnc}) expresses
  $\sigma_{j_{E\parallel}}^2$ in terms of parallel heat flux $H_\parallel$ and
  parallel and perpendicular pressures $P_\parallel$ and $P_\perp$. Dependence
  on $H_\parallel$ arises because the computational routine provided as
  Supporting Information for \citet{Gershman2015} yields uncertainties and
  covariances related to the heat flux vector $\mathbf{H}$; with this dependence
  the parallel energy flux $j_{E\parallel}$ can be written
  $j_{E\parallel} = H_\parallel + V_\parallel \left( \frac{3}{2} P_\parallel +
    P_\perp \right)$ \citep{Paschmann1998}.
  Equation (\ref{eqEFluxUnc}) also assumes (i) gyrotropy, because FAST ion and
  electron ESAs measure only one direction perpendicular to the geomagnetic
  field, and (ii) average perpendicular velocity $V_\perp = 0$, since there is
  negligible dependence on $V_\perp$ at FAST altitudes.

  Calculation of moment uncertainties and covariances from $f(\mathbf{v})$ in
  equations (\ref{eqUncerts}) requires the following assumptions:
  \begin{enumerate}

  \item The sampling of each phase space volume is unique. For FAST ESAs, which
    sample energy and pitch angle, this assumption means, for example, that
    there is no overlap between regions of phase space sampled by each
    energy-angle detector bin, and that there is no crosstalk.

  \item The sampled phase space density $f(\mathbf{v})$ corresponds to a number
    of counts $N(\mathbf{v}) = f(\mathbf{v}) \Delta V(\mathbf{v})\Delta
    X(\mathbf{v})$, where $\Delta V(\mathbf{v})$ and $\Delta X (\mathbf{v})$ are
    respectively the phase space velocity and position volumes sampled by FAST
    ESAs, and $N(\mathbf{v})$ is a Poisson-distributed random variable.

  \end{enumerate}

  The covariance between moments $\langle n A_i \rangle$ and
  $\langle n A_j \rangle$ is
  $\sigma_{\langle n A_i \rangle,\langle n A_j \rangle} = E \left [\langle n A_i
    \rangle\langle n A_j \rangle \right ] - E \left [\langle n A_i \rangle
  \right ] E \left [\langle n A_j \rangle \right ]$, where $E$ denotes the
  expectation value such that
  \begin{linenomath*}
    \begin{align}
      \begin{split}
        E \left [\langle n A_i \rangle\langle n A_j \rangle \right ] &= \iiint
        \mathbf{d}^3 \mathbf{v} A_i (\mathbf{v}) \iiint \mathbf{d}^3 \mathbf{v'}
        A_j (\mathbf{v'}) E \left [f(\mathbf{v}) f(\mathbf{v'}) \right ];
        \\
        E \left [\langle n A_i \rangle \right ] E \left [\langle n A_j \rangle \right
        ] &= \iiint \mathbf{d}^3 \mathbf{v} A_i (\mathbf{v}) \iiint \mathbf{d}^3
        \mathbf{v'} A_j (\mathbf{v'}) E \left [f(\mathbf{v}) \right ] E \left [
        f(\mathbf{v'}) \right ].
      \end{split}
    \end{align}
  \end{linenomath*}
  It follows that
  $ \sigma_{\langle n A_i \rangle,\langle n A_j \rangle } = \iiint \mathbf{d}^3
  \mathbf{v} A_i (\mathbf{v}) \iiint \mathbf{d}^3 \mathbf{v'} A_j (\mathbf{v'})
  \sigma_{f(\mathbf{v}),f(\mathbf{v'})}$; that is, the covariance between any
  two moments of $f(\mathbf{v})$ depends on the covariance between the points in
  phase space $\mathbf{v}$ and $\mathbf{v'}$. \citet{Gershman2015} show that if
  $\sigma_{f(\mathbf{v}),f(\mathbf{v'})}$ is written in terms of the correlation
  between regions of phase space,
  \begin{linenomath*}
    \begin{equation}
      \sigma_{f(\mathbf{v}),f(\mathbf{v'})} = \sigma_{f(\mathbf{v})} \sigma_{f(\mathbf{v'})}r(\mathbf{v},\mathbf{v'}),
    \end{equation}
  \end{linenomath*}
  the first assumption implies $r(\mathbf{v},\mathbf{v'}) \approx
  \delta_{\mathbf{v v'}}$, while the second assumption implies that the
  uncertainty of the sampled phase space density is $\sigma_{f(\mathbf{v})} = f
  (\mathbf{v}) \big / \sqrt{N (\mathbf{v})}$. Thus
  \begin{linenomath*}
    \begin{equation}
      \sigma_{f(\mathbf{v}),f(\mathbf{v'})} \approx
      \frac{f^2(\mathbf{v})}{N(\mathbf{v})}, 
    \end{equation}
  \end{linenomath*}
  which leads to the analytic expression
  \begin{linenomath*}
    \begin{equation} \label{eqFin} \sigma_{\langle n A_i \rangle,\langle n
        A_j\rangle } \approx \iiint \left ( \mathbf{d}^3 \mathbf{v} \right )^2 A_i
      (\mathbf{v}) A_j (\mathbf{v}) \frac{f^2(\mathbf{v})}{N(\mathbf{v})} =
      \left \langle n A_i A_j \left (\mathbf{d}^3 \mathbf{v} \right )
      \frac{f(\mathbf{v})}{N(\mathbf{v})} \right \rangle,
    \end{equation}
  \end{linenomath*}
  where the RHS of \ref{eqFin} represents $\sigma_{\langle n A_i \rangle,\langle
    n A_j\rangle }$ as a moment of $f(\mathbf{v})$.

%
%

%

%


%
%
\bibliography{Kappa2}

\begin{thebibliography}{}

\bibitem [\protect \citeauthoryear {%
Andersson%
\ \protect \BOthers {.}}{%
Andersson%
\ \protect \BOthers {.}}{%
{\protect \APACyear {2002}}%
}]{%
Andersson2002a}
\APACinsertmetastar {%
Andersson2002a}%
\begin{APACrefauthors}%
Andersson, L.%
, Ivchenko, N.%
, Clemmons, J.%
, Namgaladze, A\BPBI A.%
, Gustavsson, B.%
, Wahlund, J\BPBI E.%
\BDBL {}Yurik, R\BPBI Y.%
\end{APACrefauthors}%
\unskip\
\newblock
\APACrefYearMonthDay{2002}{}{}.
\newblock
{\BBOQ}\APACrefatitle {{Electron signatures and Alfv{\'{e}}n waves}} {{Electron
  signatures and Alfv{\'{e}}n waves}}.{\BBCQ}
\newblock
\APACjournalVolNumPages{J. Geophys. Res. Sp. Phys.}{107}{A9}{1244}.
\newblock
\begin{APACrefDOI} \doi{10.1029/2001JA900096} \end{APACrefDOI}
\PrintBackRefs{\CurrentBib}

\bibitem [\protect \citeauthoryear {%
Bostr{\"{o}}m%
}{%
Bostr{\"{o}}m%
}{%
{\protect \APACyear {2003}}%
}]{%
Bostrom2003a}
\APACinsertmetastar {%
Bostrom2003a}%
\begin{APACrefauthors}%
Bostr{\"{o}}m, R.%
\end{APACrefauthors}%
\unskip\
\newblock
\APACrefYearMonthDay{2003}{apr}{}.
\newblock
{\BBOQ}\APACrefatitle {{Kinetic and space charge control of current flow and
  voltage drops along magnetic flux tubes: Kinetic effects}} {{Kinetic and
  space charge control of current flow and voltage drops along magnetic flux
  tubes: Kinetic effects}}.{\BBCQ}
\newblock
\APACjournalVolNumPages{J. Geophys. Res. Sp. Phys.}{108}{A4}{}.
\newblock
\begin{APACrefURL} \url{http://dx.doi.org/10.1029/2002JA009295}
  \end{APACrefURL}
\newblock
\begin{APACrefDOI} \doi{10.1029/2002JA009295} \end{APACrefDOI}
\PrintBackRefs{\CurrentBib}

\bibitem [\protect \citeauthoryear {%
Bostr{\"{o}}m%
}{%
Bostr{\"{o}}m%
}{%
{\protect \APACyear {2004}}%
}]{%
Bostrom2004}
\APACinsertmetastar {%
Bostrom2004}%
\begin{APACrefauthors}%
Bostr{\"{o}}m, R.%
\end{APACrefauthors}%
\unskip\
\newblock
\APACrefYearMonthDay{2004}{}{}.
\newblock
{\BBOQ}\APACrefatitle {{Kinetic and space charge control of current flow and
  voltage drops along magnetic flux tubes: 2. Space charge effects}} {{Kinetic
  and space charge control of current flow and voltage drops along magnetic
  flux tubes: 2. Space charge effects}}.{\BBCQ}
\newblock
\APACjournalVolNumPages{J. Geophys. Res. Sp. Phys.}{109}{A1}{}.
\newblock
\begin{APACrefURL} \url{http://dx.doi.org/10.1029/2003JA010078}
  \end{APACrefURL}
\newblock
\begin{APACrefDOI} \doi{10.1029/2003JA010078} \end{APACrefDOI}
\PrintBackRefs{\CurrentBib}

\bibitem [\protect \citeauthoryear {%
Carlson%
, Mcfadden%
, Turin%
, Curtis%
\BCBL {}\ \BBA {} Magoncelli%
}{%
Carlson%
\ \protect \BOthers {.}}{%
{\protect \APACyear {2001}}%
}]{%
Carlson2001}
\APACinsertmetastar {%
Carlson2001}%
\begin{APACrefauthors}%
Carlson, C\BPBI W.%
, Mcfadden, J\BPBI P.%
, Turin, P.%
, Curtis, D\BPBI W.%
\BCBL {}\ \BBA {} Magoncelli, A.%
\end{APACrefauthors}%
\unskip\
\newblock
\APACrefYearMonthDay{2001}{}{}.
\newblock
{\BBOQ}\APACrefatitle {{The electron and ion plasma experiment for FAST}} {{The
  electron and ion plasma experiment for FAST}}.{\BBCQ}
\newblock
\APACjournalVolNumPages{Space Sci. Rev.}{98}{1}{33--66}.
\newblock
\begin{APACrefURL} \url{http://dx.doi.org/10.1023/A:1013139910140}
  \end{APACrefURL}
\newblock
\begin{APACrefDOI} \doi{10.1023/A:1013139910140} \end{APACrefDOI}
\PrintBackRefs{\CurrentBib}

\bibitem [\protect \citeauthoryear {%
Chiu%
\ \BBA {} Schulz%
}{%
Chiu%
\ \BBA {} Schulz%
}{%
{\protect \APACyear {1978}}%
}]{%
Chiu1978}
\APACinsertmetastar {%
Chiu1978}%
\begin{APACrefauthors}%
Chiu, Y\BPBI T.%
\BCBT {}\ \BBA {} Schulz, M.%
\end{APACrefauthors}%
\unskip\
\newblock
\APACrefYearMonthDay{1978}{feb}{}.
\newblock
{\BBOQ}\APACrefatitle {{Self-consistent particle and parallel electrostatic
  field distributions in the magnetospheric-ionospheric auroral region}}
  {{Self-consistent particle and parallel electrostatic field distributions in
  the magnetospheric-ionospheric auroral region}}.{\BBCQ}
\newblock
\APACjournalVolNumPages{Journal of Geophysical Research: Space
  Physics}{83}{A2}{629--642}.
\newblock
\begin{APACrefURL} \url{http://dx.doi.org/10.1029/JA083iA02p00629}
  \end{APACrefURL}
\newblock
\begin{APACrefDOI} \doi{10.1029/JA083iA02p00629} \end{APACrefDOI}
\PrintBackRefs{\CurrentBib}

\bibitem [\protect \citeauthoryear {%
Christon%
, Williams%
, Mitchell%
, Frank%
\BCBL {}\ \BBA {} Huang%
}{%
Christon%
\ \protect \BOthers {.}}{%
{\protect \APACyear {1989}}%
}]{%
Christon1989}
\APACinsertmetastar {%
Christon1989}%
\begin{APACrefauthors}%
Christon, S\BPBI P.%
, Williams, D\BPBI J.%
, Mitchell, D\BPBI G.%
, Frank, L\BPBI A.%
\BCBL {}\ \BBA {} Huang, C\BPBI Y.%
\end{APACrefauthors}%
\unskip\
\newblock
\APACrefYearMonthDay{1989}{}{}.
\newblock
{\BBOQ}\APACrefatitle {{Spectral characteristics of plasma sheet ion and
  electron populations during undisturbed geomagnetic conditions}} {{Spectral
  characteristics of plasma sheet ion and electron populations during
  undisturbed geomagnetic conditions}}.{\BBCQ}
\newblock
\APACjournalVolNumPages{Journal of Geophysical Research}{94}{A10}{13409}.
\newblock
\begin{APACrefURL} \url{http://doi.wiley.com/10.1029/JA094iA10p13409}
  \end{APACrefURL}
\newblock
\begin{APACrefDOI} \doi{10.1029/JA094iA10p13409} \end{APACrefDOI}
\PrintBackRefs{\CurrentBib}

\bibitem [\protect \citeauthoryear {%
Christon%
, Williams%
, Mitchell%
, Huang%
\BCBL {}\ \BBA {} Frank%
}{%
Christon%
\ \protect \BOthers {.}}{%
{\protect \APACyear {1991}}%
}]{%
Christon1991}
\APACinsertmetastar {%
Christon1991}%
\begin{APACrefauthors}%
Christon, S\BPBI P.%
, Williams, D\BPBI J.%
, Mitchell, D\BPBI G.%
, Huang, C\BPBI Y.%
\BCBL {}\ \BBA {} Frank, L\BPBI A.%
\end{APACrefauthors}%
\unskip\
\newblock
\APACrefYearMonthDay{1991}{}{}.
\newblock
{\BBOQ}\APACrefatitle {{Spectral characteristics of plasma sheet ion and
  electron populations during disturbed geomagnetic conditions}} {{Spectral
  characteristics of plasma sheet ion and electron populations during disturbed
  geomagnetic conditions}}.{\BBCQ}
\newblock
\APACjournalVolNumPages{Journal of Geophysical Research}{96}{A1}{1}.
\newblock
\begin{APACrefURL} \url{http://doi.wiley.com/10.1029/90JA01633}
  \end{APACrefURL}
\newblock
\begin{APACrefDOI} \doi{10.1029/90JA01633} \end{APACrefDOI}
\PrintBackRefs{\CurrentBib}

\bibitem [\protect \citeauthoryear {%
Cowley%
}{%
Cowley%
}{%
{\protect \APACyear {2000}}%
}]{%
Cowley2000}
\APACinsertmetastar {%
Cowley2000}%
\begin{APACrefauthors}%
Cowley, S\BPBI W\BPBI H.%
\end{APACrefauthors}%
\unskip\
\newblock
\APACrefYearMonthDay{2000}{}{}.
\newblock
{\BBOQ}\APACrefatitle {{Magnetosphere-Ionosphere Interactions: A Tutorial
  Review}} {{Magnetosphere-Ionosphere Interactions: A Tutorial Review}}.{\BBCQ}
\newblock
\BIn{} S\BHBI I.~Ohtani, R.~Fujii, M.~Hesse\BCBL {}\ \BBA {} R\BPBI L.~Lysak\
  (\BEDS), \APACrefbtitle {Magnetos. Curr. Syst.} {Magnetos. curr. syst.}\
  (\BVOL~118, \BPGS\ 91--106).
\newblock
\APACaddressPublisher{Washington, D.C.}{American Geophysical Union}.
\newblock
\begin{APACrefURL} \url{http://dx.doi.org/10.1029/GM118p0091} \end{APACrefURL}
\newblock
\begin{APACrefDOI} \doi{10.1029/GM118p0091} \end{APACrefDOI}
\PrintBackRefs{\CurrentBib}

\bibitem [\protect \citeauthoryear {%
Dombeck%
, Cattell%
\BCBL {}\ \BBA {} McFadden%
}{%
Dombeck%
\ \protect \BOthers {.}}{%
{\protect \APACyear {2013}}%
}]{%
Dombeck2013}
\APACinsertmetastar {%
Dombeck2013}%
\begin{APACrefauthors}%
Dombeck, J.%
, Cattell, C.%
\BCBL {}\ \BBA {} McFadden, J.%
\end{APACrefauthors}%
\unskip\
\newblock
\APACrefYearMonthDay{2013}{sep}{}.
\newblock
{\BBOQ}\APACrefatitle {{A FAST study of quasi-static structure ("Inverted-V")
  potential drops and their latitudinal dependence in the premidnight sector
  and ramifications for the current-voltage relationship}} {{A FAST study of
  quasi-static structure ("Inverted-V") potential drops and their latitudinal
  dependence in the premidnight sector and ramifications for the
  current-voltage relationship}}.{\BBCQ}
\newblock
\APACjournalVolNumPages{J. Geophys. Res. Sp. Phys.}{118}{9}{5731--5741}.
\newblock
\begin{APACrefURL} \url{http://dx.doi.org/10.1002/jgra.50532} \end{APACrefURL}
\newblock
\begin{APACrefDOI} \doi{10.1002/jgra.50532} \end{APACrefDOI}
\PrintBackRefs{\CurrentBib}

\bibitem [\protect \citeauthoryear {%
Dors%
\ \BBA {} Kletzing%
}{%
Dors%
\ \BBA {} Kletzing%
}{%
{\protect \APACyear {1999}}%
}]{%
Dors1999}
\APACinsertmetastar {%
Dors1999}%
\begin{APACrefauthors}%
Dors, E\BPBI E.%
\BCBT {}\ \BBA {} Kletzing, C\BPBI A.%
\end{APACrefauthors}%
\unskip\
\newblock
\APACrefYearMonthDay{1999}{apr}{}.
\newblock
{\BBOQ}\APACrefatitle {{Effects of suprathermal tails on auroral
  electrodynamics}} {{Effects of suprathermal tails on auroral
  electrodynamics}}.{\BBCQ}
\newblock
\APACjournalVolNumPages{Journal of Geophysical Research: Space
  Physics}{104}{A4}{6783--6796}.
\newblock
\begin{APACrefURL} \url{http://doi.wiley.com/10.1029/1998JA900135}
  \end{APACrefURL}
\newblock
\begin{APACrefDOI} \doi{10.1029/1998JA900135} \end{APACrefDOI}
\PrintBackRefs{\CurrentBib}

\bibitem [\protect \citeauthoryear {%
Elphic%
\ \protect \BOthers {.}}{%
Elphic%
\ \protect \BOthers {.}}{%
{\protect \APACyear {1998}}%
}]{%
Elphic1998}
\APACinsertmetastar {%
Elphic1998}%
\begin{APACrefauthors}%
Elphic, R\BPBI C.%
, Bonnell, J\BPBI W.%
, Strangeway, R\BPBI J.%
, Kepko, L.%
, Ergun, R\BPBI E.%
, McFadden, J\BPBI P.%
\BDBL {}Pfaff, R.%
\end{APACrefauthors}%
\unskip\
\newblock
\APACrefYearMonthDay{1998}{jun}{}.
\newblock
{\BBOQ}\APACrefatitle {{The auroral current circuit and field-aligned currents
  observed by FAST}} {{The auroral current circuit and field-aligned currents
  observed by FAST}}.{\BBCQ}
\newblock
\APACjournalVolNumPages{Geophysical Research Letters}{25}{12}{2033--2036}.
\newblock
\begin{APACrefURL} \url{http://doi.wiley.com/10.1029/98GL01158}
  \end{APACrefURL}
\newblock
\begin{APACrefDOI} \doi{10.1029/98GL01158} \end{APACrefDOI}
\PrintBackRefs{\CurrentBib}

\bibitem [\protect \citeauthoryear {%
Forsyth%
\ \protect \BOthers {.}}{%
Forsyth%
\ \protect \BOthers {.}}{%
{\protect \APACyear {2012}}%
}]{%
Forsyth2012}
\APACinsertmetastar {%
Forsyth2012}%
\begin{APACrefauthors}%
Forsyth, C.%
, Fazakerley, A\BPBI N.%
, Walsh, A\BPBI P.%
, Watt, C\BPBI E\BPBI J.%
, Garza, K\BPBI J.%
, Owen, C\BPBI J.%
\BDBL {}Doss, N.%
\end{APACrefauthors}%
\unskip\
\newblock
\APACrefYearMonthDay{2012}{dec}{}.
\newblock
{\BBOQ}\APACrefatitle {{Temporal evolution and electric potential structure of
  the auroral acceleration region from multispacecraft measurements}}
  {{Temporal evolution and electric potential structure of the auroral
  acceleration region from multispacecraft measurements}}.{\BBCQ}
\newblock
\APACjournalVolNumPages{Journal of Geophysical Research: Space
  Physics}{117}{A12}{}.
\newblock
\begin{APACrefURL} \url{http://dx.doi.org/10.1029/2012JA017655}
  \end{APACrefURL}
\newblock
\begin{APACrefDOI} \doi{10.1029/2012JA017655} \end{APACrefDOI}
\PrintBackRefs{\CurrentBib}

\bibitem [\protect \citeauthoryear {%
Gershman%
, Dorelli%
, F.-Vi{\~{n}}as%
\BCBL {}\ \BBA {} Pollock%
}{%
Gershman%
\ \protect \BOthers {.}}{%
{\protect \APACyear {2015}}%
}]{%
Gershman2015}
\APACinsertmetastar {%
Gershman2015}%
\begin{APACrefauthors}%
Gershman, D\BPBI J.%
, Dorelli, J\BPBI C.%
, F.-Vi{\~{n}}as, A.%
\BCBL {}\ \BBA {} Pollock, C\BPBI J.%
\end{APACrefauthors}%
\unskip\
\newblock
\APACrefYearMonthDay{2015}{aug}{}.
\newblock
{\BBOQ}\APACrefatitle {{The calculation of moment uncertainties from velocity
  distribution functions with random errors}} {{The calculation of moment
  uncertainties from velocity distribution functions with random
  errors}}.{\BBCQ}
\newblock
\APACjournalVolNumPages{Journal of Geophysical Research: Space
  Physics}{120}{8}{6633--6645}.
\newblock
\begin{APACrefURL} \url{http://dx.doi.org/10.1002/2014JA020775}
  \end{APACrefURL}
\newblock
\begin{APACrefDOI} \doi{10.1002/2014JA020775} \end{APACrefDOI}
\PrintBackRefs{\CurrentBib}

\bibitem [\protect \citeauthoryear {%
Hatch%
, Chaston%
\BCBL {}\ \BBA {} LaBelle%
}{%
Hatch%
\ \protect \BOthers {.}}{%
{\protect \APACyear {2018}}%
}]{%
Hatch2018}
\APACinsertmetastar {%
Hatch2018}%
\begin{APACrefauthors}%
Hatch, S\BPBI M.%
, Chaston, C\BPBI C.%
\BCBL {}\ \BBA {} LaBelle, J.%
\end{APACrefauthors}%
\unskip\
\newblock
\APACrefYearMonthDay{2018}{aug}{}.
\newblock
{\BBOQ}\APACrefatitle {{Nonthermal limit of monoenergetic precipitation in the
  auroral acceleration region}} {{Nonthermal limit of monoenergetic
  precipitation in the auroral acceleration region}}.{\BBCQ}
\newblock
\APACjournalVolNumPages{Geophysical Research Letters}{45}{19}{10167--10176}.
\newblock
\begin{APACrefURL} \url{https://doi.org/10.1029/2018GL078948
  http://doi.wiley.com/10.1029/2018GL078948} \end{APACrefURL}
\newblock
\begin{APACrefDOI} \doi{10.1029/2018GL078948} \end{APACrefDOI}
\PrintBackRefs{\CurrentBib}

\bibitem [\protect \citeauthoryear {%
Hull%
\ \protect \BOthers {.}}{%
Hull%
\ \protect \BOthers {.}}{%
{\protect \APACyear {2010}}%
}]{%
Hull2010}
\APACinsertmetastar {%
Hull2010}%
\begin{APACrefauthors}%
Hull, A\BPBI J.%
, Wilber, M.%
, Chaston, C\BPBI C.%
, Bonnell, J\BPBI W.%
, McFadden, J\BPBI P.%
, Mozer, F\BPBI S.%
\BDBL {}Goldstein, M\BPBI L.%
\end{APACrefauthors}%
\unskip\
\newblock
\APACrefYearMonthDay{2010}{jun}{}.
\newblock
{\BBOQ}\APACrefatitle {{Time development of field-aligned currents, potential
  drops, and plasma associated with an auroral poleward boundary
  intensification}} {{Time development of field-aligned currents, potential
  drops, and plasma associated with an auroral poleward boundary
  intensification}}.{\BBCQ}
\newblock
\APACjournalVolNumPages{Journal of Geophysical Research: Space
  Physics}{115}{A6}{}.
\newblock
\begin{APACrefURL} \url{http://doi.wiley.com/10.1029/2009JA014651}
  \end{APACrefURL}
\newblock
\begin{APACrefDOI} \doi{10.1029/2009JA014651} \end{APACrefDOI}
\PrintBackRefs{\CurrentBib}

\bibitem [\protect \citeauthoryear {%
Janhunen%
\ \BBA {} Olsson%
}{%
Janhunen%
\ \BBA {} Olsson%
}{%
{\protect \APACyear {1998}}%
}]{%
Janhunen1998}
\APACinsertmetastar {%
Janhunen1998}%
\begin{APACrefauthors}%
Janhunen, P.%
\BCBT {}\ \BBA {} Olsson, A.%
\end{APACrefauthors}%
\unskip\
\newblock
\APACrefYearMonthDay{1998}{}{}.
\newblock
{\BBOQ}\APACrefatitle {{The current-voltage relationship revisited: exact and
  approximate formulas with almost general validity for hot magnetospheric
  electrons for bi-Maxwellian and kappa distributions}} {{The current-voltage
  relationship revisited: exact and approximate formulas with almost general
  validity for hot magnetospheric electrons for bi-Maxwellian and kappa
  distributions}}.{\BBCQ}
\newblock
\APACjournalVolNumPages{Ann. Geophys.}{16}{3}{292--297}.
\newblock
\begin{APACrefURL} \url{http://www.ann-geophys.net/16/292/1998/}
  \end{APACrefURL}
\newblock
\begin{APACrefDOI} \doi{10.1007/s00585-998-0292-6} \end{APACrefDOI}
\PrintBackRefs{\CurrentBib}

\bibitem [\protect \citeauthoryear {%
Kaeppler%
, Nicolls%
, Str{\o}mme%
, Kletzing%
\BCBL {}\ \BBA {} Bounds%
}{%
Kaeppler%
\ \protect \BOthers {.}}{%
{\protect \APACyear {2014}}%
}]{%
Kaeppler2014a}
\APACinsertmetastar {%
Kaeppler2014a}%
\begin{APACrefauthors}%
Kaeppler, S\BPBI R.%
, Nicolls, M\BPBI J.%
, Str{\o}mme, A.%
, Kletzing, C\BPBI A.%
\BCBL {}\ \BBA {} Bounds, S\BPBI R.%
\end{APACrefauthors}%
\unskip\
\newblock
\APACrefYearMonthDay{2014}{dec}{}.
\newblock
{\BBOQ}\APACrefatitle {{Observations in the E region ionosphere of kappa
  distribution functions associated with precipitating auroral electrons and
  discrete aurorae}} {{Observations in the E region ionosphere of kappa
  distribution functions associated with precipitating auroral electrons and
  discrete aurorae}}.{\BBCQ}
\newblock
\APACjournalVolNumPages{J. Geophys. Res. Sp. Phys.}{119}{12}{10,110--164,183}.
\newblock
\begin{APACrefURL} \url{http://dx.doi.org/10.1002/2014JA020356}
  \end{APACrefURL}
\newblock
\begin{APACrefDOI} \doi{10.1002/2014JA020356} \end{APACrefDOI}
\PrintBackRefs{\CurrentBib}

\bibitem [\protect \citeauthoryear {%
Karlsson%
}{%
Karlsson%
}{%
{\protect \APACyear {2012}}%
}]{%
Karlsson2012}
\APACinsertmetastar {%
Karlsson2012}%
\begin{APACrefauthors}%
Karlsson, T.%
\end{APACrefauthors}%
\unskip\
\newblock
\APACrefYearMonthDay{2012}{}{}.
\newblock
{\BBOQ}\APACrefatitle {{The Acceleration Region of Stable Auroral Arcs}} {{The
  Acceleration Region of Stable Auroral Arcs}}.{\BBCQ}
\newblock
\BIn{} \APACrefbtitle {Auror. Phenomenol. Magnetos. Process. Earth Other
  Planets} {Auror. phenomenol. magnetos. process. earth other planets}\ (\BPGS\
  227--240).
\newblock
\APACaddressPublisher{}{American Geophysical Union}.
\newblock
\begin{APACrefURL} \url{http://dx.doi.org/10.1029/2011GM001179}
  \end{APACrefURL}
\newblock
\begin{APACrefDOI} \doi{10.1029/2011GM001179} \end{APACrefDOI}
\PrintBackRefs{\CurrentBib}

\bibitem [\protect \citeauthoryear {%
Kletzing%
, Scudder%
, Dors%
\BCBL {}\ \BBA {} Curto%
}{%
Kletzing%
\ \protect \BOthers {.}}{%
{\protect \APACyear {2003}}%
}]{%
Kletzing2003}
\APACinsertmetastar {%
Kletzing2003}%
\begin{APACrefauthors}%
Kletzing, C\BPBI A.%
, Scudder, J\BPBI D.%
, Dors, E\BPBI E.%
\BCBL {}\ \BBA {} Curto, C.%
\end{APACrefauthors}%
\unskip\
\newblock
\APACrefYearMonthDay{2003}{oct}{}.
\newblock
{\BBOQ}\APACrefatitle {{Auroral source region: Plasma properties of the
  high-latitude plasma sheet}} {{Auroral source region: Plasma properties of
  the high-latitude plasma sheet}}.{\BBCQ}
\newblock
\APACjournalVolNumPages{Journal of Geophysical Research: Space
  Physics}{108}{A10}{}.
\newblock
\begin{APACrefURL} \url{http://dx.doi.org/10.1029/2002JA009678}
  \end{APACrefURL}
\newblock
\begin{APACrefDOI} \doi{10.1029/2002JA009678} \end{APACrefDOI}
\PrintBackRefs{\CurrentBib}

\bibitem [\protect \citeauthoryear {%
Knight%
}{%
Knight%
}{%
{\protect \APACyear {1973}}%
}]{%
Knight1973}
\APACinsertmetastar {%
Knight1973}%
\begin{APACrefauthors}%
Knight, S.%
\end{APACrefauthors}%
\unskip\
\newblock
\APACrefYearMonthDay{1973}{may}{}.
\newblock
{\BBOQ}\APACrefatitle {{Parallel electric fields}} {{Parallel electric
  fields}}.{\BBCQ}
\newblock
\APACjournalVolNumPages{Planet. Space Sci.}{21}{}{741--750}.
\newblock
\begin{APACrefDOI} \doi{10.1016/0032-0633(73)90093-7} \end{APACrefDOI}
\PrintBackRefs{\CurrentBib}

\bibitem [\protect \citeauthoryear {%
Li%
\ \protect \BOthers {.}}{%
Li%
\ \protect \BOthers {.}}{%
{\protect \APACyear {2014}}%
}]{%
Li2014}
\APACinsertmetastar {%
Li2014}%
\begin{APACrefauthors}%
Li, B.%
, Marklund, G.%
, Alm, L.%
, Karlsson, T.%
, Lindqvist, P\BHBI A.%
\BCBL {}\ \BBA {} Masson, A.%
\end{APACrefauthors}%
\unskip\
\newblock
\APACrefYearMonthDay{2014}{nov}{}.
\newblock
{\BBOQ}\APACrefatitle {{Statistical altitude distribution of Cluster auroral
  electric fields, indicating mainly quasi-static acceleration below 2.8 RE and
  Alfv{\'{e}}nic above}} {{Statistical altitude distribution of Cluster auroral
  electric fields, indicating mainly quasi-static acceleration below 2.8 RE and
  Alfv{\'{e}}nic above}}.{\BBCQ}
\newblock
\APACjournalVolNumPages{Journal of Geophysical Research: Space
  Physics}{119}{11}{8984--8991}.
\newblock
\begin{APACrefURL} \url{http://dx.doi.org/10.1002/2014JA020225}
  \end{APACrefURL}
\newblock
\begin{APACrefDOI} \doi{10.1002/2014JA020225} \end{APACrefDOI}
\PrintBackRefs{\CurrentBib}

\bibitem [\protect \citeauthoryear {%
Liemohn%
\ \BBA {} Khazanov%
}{%
Liemohn%
\ \BBA {} Khazanov%
}{%
{\protect \APACyear {1998}}%
}]{%
Liemohn1998}
\APACinsertmetastar {%
Liemohn1998}%
\begin{APACrefauthors}%
Liemohn, M\BPBI W.%
\BCBT {}\ \BBA {} Khazanov, G\BPBI V.%
\end{APACrefauthors}%
\unskip\
\newblock
\APACrefYearMonthDay{1998}{mar}{}.
\newblock
{\BBOQ}\APACrefatitle {{Collisionless plasma modeling in an arbitrary potential
  energy distribution}} {{Collisionless plasma modeling in an arbitrary
  potential energy distribution}}.{\BBCQ}
\newblock
\APACjournalVolNumPages{Physics of Plasmas}{5}{3}{580--589}.
\newblock
\begin{APACrefURL} \url{https://doi.org/10.1063/1.872750} \end{APACrefURL}
\newblock
\begin{APACrefDOI} \doi{10.1063/1.872750} \end{APACrefDOI}
\PrintBackRefs{\CurrentBib}

\bibitem [\protect \citeauthoryear {%
Livadiotis%
\ \BBA {} McComas%
}{%
Livadiotis%
\ \BBA {} McComas%
}{%
{\protect \APACyear {2010}}%
}]{%
Livadiotis2010}
\APACinsertmetastar {%
Livadiotis2010}%
\begin{APACrefauthors}%
Livadiotis, G.%
\BCBT {}\ \BBA {} McComas, D\BPBI J.%
\end{APACrefauthors}%
\unskip\
\newblock
\APACrefYearMonthDay{2010}{may}{}.
\newblock
{\BBOQ}\APACrefatitle {{Exploring transitions of space plasmas out of
  equilibrium}} {{Exploring transitions of space plasmas out of
  equilibrium}}.{\BBCQ}
\newblock
\APACjournalVolNumPages{The Astrophysical Journal}{714}{1}{971--987}.
\newblock
\begin{APACrefURL}
  \url{http://stacks.iop.org/0004-637X/714/i=1/a=971?key=crossref.21d2f7d6f92c9dede5d5fca69bb11fe1}
  \end{APACrefURL}
\newblock
\begin{APACrefDOI} \doi{10.1088/0004-637X/714/1/971} \end{APACrefDOI}
\PrintBackRefs{\CurrentBib}

\bibitem [\protect \citeauthoryear {%
Lu%
, Reiff%
, Burch%
\BCBL {}\ \BBA {} Winningham%
}{%
Lu%
\ \protect \BOthers {.}}{%
{\protect \APACyear {1991}}%
}]{%
Lu1991}
\APACinsertmetastar {%
Lu1991}%
\begin{APACrefauthors}%
Lu, G.%
, Reiff, P\BPBI H.%
, Burch, J\BPBI L.%
\BCBL {}\ \BBA {} Winningham, J\BPBI D.%
\end{APACrefauthors}%
\unskip\
\newblock
\APACrefYearMonthDay{1991}{aug}{}.
\newblock
{\BBOQ}\APACrefatitle {{On the auroral current-voltage relationship}} {{On the
  auroral current-voltage relationship}}.{\BBCQ}
\newblock
\APACjournalVolNumPages{Journal of Geophysical Research}{96}{A3}{3523}.
\newblock
\begin{APACrefURL} \url{https://doi.org/10.1029/90JA02462
  http://doi.wiley.com/10.1029/90JA02462} \end{APACrefURL}
\newblock
\begin{APACrefDOI} \doi{10.1029/90JA02462} \end{APACrefDOI}
\PrintBackRefs{\CurrentBib}

\bibitem [\protect \citeauthoryear {%
Marghitu%
, Klecker%
\BCBL {}\ \BBA {} McFadden%
}{%
Marghitu%
\ \protect \BOthers {.}}{%
{\protect \APACyear {2006}}%
}]{%
Marghitu2006}
\APACinsertmetastar {%
Marghitu2006}%
\begin{APACrefauthors}%
Marghitu, O.%
, Klecker, B.%
\BCBL {}\ \BBA {} McFadden, J\BPBI P.%
\end{APACrefauthors}%
\unskip\
\newblock
\APACrefYearMonthDay{2006}{}{}.
\newblock
{\BBOQ}\APACrefatitle {{The anisotropy of precipitating auroral electrons: A
  FAST case study}} {{The anisotropy of precipitating auroral electrons: A FAST
  case study}}.{\BBCQ}
\newblock
\APACjournalVolNumPages{Advances in Space Research}{38}{8}{1694--1701}.
\newblock
\begin{APACrefURL}
  \url{http://www.sciencedirect.com/science/article/pii/S0273117706001529}
  \end{APACrefURL}
\newblock
\begin{APACrefDOI} \doi{https://doi.org/10.1016/j.asr.2006.03.028}
  \end{APACrefDOI}
\PrintBackRefs{\CurrentBib}

\bibitem [\protect \citeauthoryear {%
Marklund%
\ \protect \BOthers {.}}{%
Marklund%
\ \protect \BOthers {.}}{%
{\protect \APACyear {2011}}%
}]{%
Marklund2011}
\APACinsertmetastar {%
Marklund2011}%
\begin{APACrefauthors}%
Marklund, G\BPBI T.%
, Sadeghi, S.%
, Karlsson, T.%
, Lindqvist, P\BHBI A.%
, Nilsson, H.%
, Forsyth, C.%
\BDBL {}Pickett, J.%
\end{APACrefauthors}%
\unskip\
\newblock
\APACrefYearMonthDay{2011}{feb}{}.
\newblock
{\BBOQ}\APACrefatitle {{Altitude Distribution of the Auroral Acceleration
  Potential Determined from Cluster Satellite Data at Different Heights}}
  {{Altitude Distribution of the Auroral Acceleration Potential Determined from
  Cluster Satellite Data at Different Heights}}.{\BBCQ}
\newblock
\APACjournalVolNumPages{Physical Review Letters}{106}{5}{55002}.
\newblock
\begin{APACrefURL}
  \url{https://link.aps.org/doi/10.1103/PhysRevLett.106.055002}
  \end{APACrefURL}
\newblock
\begin{APACrefDOI} \doi{10.1103/PhysRevLett.106.055002} \end{APACrefDOI}
\PrintBackRefs{\CurrentBib}

\bibitem [\protect \citeauthoryear {%
Moore%
, Pollock%
\BCBL {}\ \BBA {} Young%
}{%
Moore%
\ \protect \BOthers {.}}{%
{\protect \APACyear {1998}}%
}]{%
Moore1998}
\APACinsertmetastar {%
Moore1998}%
\begin{APACrefauthors}%
Moore, T\BPBI E.%
, Pollock, C\BPBI J.%
\BCBL {}\ \BBA {} Young, D\BPBI T.%
\end{APACrefauthors}%
\unskip\
\newblock
\APACrefYearMonthDay{1998}{}{}.
\newblock
{\BBOQ}\APACrefatitle {{Kinetic Core Plasma Diagnostics}} {{Kinetic Core Plasma
  Diagnostics}}.{\BBCQ}
\newblock
\BIn{} \APACrefbtitle {Measurement Techniques in Space Plasmas:Particles}
  {Measurement techniques in space plasmas:particles}\ (\BPGS\ 105--123).
\newblock
\APACaddressPublisher{}{American Geophysical Union}.
\newblock
\begin{APACrefURL} \url{http://dx.doi.org/10.1029/GM102p0105} \end{APACrefURL}
\newblock
\begin{APACrefDOI} \doi{10.1029/GM102p0105} \end{APACrefDOI}
\PrintBackRefs{\CurrentBib}

\bibitem [\protect \citeauthoryear {%
Morooka%
, Mukai%
\BCBL {}\ \BBA {} Fukunishi%
}{%
Morooka%
\ \protect \BOthers {.}}{%
{\protect \APACyear {2004}}%
}]{%
Morooka2004}
\APACinsertmetastar {%
Morooka2004}%
\begin{APACrefauthors}%
Morooka, M.%
, Mukai, T.%
\BCBL {}\ \BBA {} Fukunishi, H.%
\end{APACrefauthors}%
\unskip\
\newblock
\APACrefYearMonthDay{2004}{}{}.
\newblock
{\BBOQ}\APACrefatitle {{Current-voltage relationship in the auroral particle
  acceleration region}} {{Current-voltage relationship in the auroral particle
  acceleration region}}.{\BBCQ}
\newblock
\APACjournalVolNumPages{Annales Geophysicae}{22}{10}{3641--3655}.
\newblock
\begin{APACrefURL} \url{http://www.ann-geophys.net/22/3641/2004/}
  \end{APACrefURL}
\newblock
\begin{APACrefDOI} \doi{10.5194/angeo-22-3641-2004} \end{APACrefDOI}
\PrintBackRefs{\CurrentBib}

\bibitem [\protect \citeauthoryear {%
Mozer%
\ \BBA {} Hull%
}{%
Mozer%
\ \BBA {} Hull%
}{%
{\protect \APACyear {2001}}%
}]{%
Mozer2001}
\APACinsertmetastar {%
Mozer2001}%
\begin{APACrefauthors}%
Mozer, F\BPBI S.%
\BCBT {}\ \BBA {} Hull, A.%
\end{APACrefauthors}%
\unskip\
\newblock
\APACrefYearMonthDay{2001}{apr}{}.
\newblock
{\BBOQ}\APACrefatitle {{Origin and geometry of upward parallel electric fields
  in the auroral acceleration region}} {{Origin and geometry of upward parallel
  electric fields in the auroral acceleration region}}.{\BBCQ}
\newblock
\APACjournalVolNumPages{Journal of Geophysical Research: Space
  Physics}{106}{A4}{5763--5778}.
\newblock
\begin{APACrefURL} \url{http://dx.doi.org/10.1029/2000JA900117}
  \end{APACrefURL}
\newblock
\begin{APACrefDOI} \doi{10.1029/2000JA900117} \end{APACrefDOI}
\PrintBackRefs{\CurrentBib}

\bibitem [\protect \citeauthoryear {%
Nicholls%
, Dopita%
\BCBL {}\ \BBA {} Sutherland%
}{%
Nicholls%
\ \protect \BOthers {.}}{%
{\protect \APACyear {2012}}%
}]{%
Nicholls2012}
\APACinsertmetastar {%
Nicholls2012}%
\begin{APACrefauthors}%
Nicholls, D\BPBI C.%
, Dopita, M\BPBI A.%
\BCBL {}\ \BBA {} Sutherland, R\BPBI S.%
\end{APACrefauthors}%
\unskip\
\newblock
\APACrefYearMonthDay{2012}{jun}{}.
\newblock
{\BBOQ}\APACrefatitle {{RESOLVING THE ELECTRON TEMPERATURE DISCREPANCIES IN H
  II REGIONS AND PLANETARY NEBULAE: $\kappa$-DISTRIBUTED ELECTRONS}}
  {{RESOLVING THE ELECTRON TEMPERATURE DISCREPANCIES IN H II REGIONS AND
  PLANETARY NEBULAE: $\kappa$-DISTRIBUTED ELECTRONS}}.{\BBCQ}
\newblock
\APACjournalVolNumPages{The Astrophysical Journal}{752}{2}{148}.
\newblock
\begin{APACrefURL} \url{http://stacks.iop.org/0004-637X/752/i=2/a=148
  http://stacks.iop.org/0004-637X/752/i=2/a=148?key=crossref.ebaf4926cac189b27a01f14f79469281}
  \end{APACrefURL}
\newblock
\begin{APACrefDOI} \doi{10.1088/0004-637X/752/2/148} \end{APACrefDOI}
\PrintBackRefs{\CurrentBib}

\bibitem [\protect \citeauthoryear {%
Ogasawara%
\ \protect \BOthers {.}}{%
Ogasawara%
\ \protect \BOthers {.}}{%
{\protect \APACyear {2017}}%
}]{%
Ogasawara2017}
\APACinsertmetastar {%
Ogasawara2017}%
\begin{APACrefauthors}%
Ogasawara, K.%
, Livadiotis, G.%
, Grubbs, G\BPBI A.%
, Jahn, J\BHBI M.%
, Michell, R.%
, Samara, M.%
\BDBL {}Winningham, J\BPBI D.%
\end{APACrefauthors}%
\unskip\
\newblock
\APACrefYearMonthDay{2017}{apr}{}.
\newblock
{\BBOQ}\APACrefatitle {{Properties of suprathermal electrons associated with
  discrete auroral arcs}} {{Properties of suprathermal electrons associated
  with discrete auroral arcs}}.{\BBCQ}
\newblock
\APACjournalVolNumPages{Geophysical Research Letters}{44}{8}{3475--3484}.
\newblock
\begin{APACrefURL} \url{http://dx.doi.org/10.1002/2017GL072715}
  \end{APACrefURL}
\newblock
\begin{APACrefDOI} \doi{10.1002/2017GL072715} \end{APACrefDOI}
\PrintBackRefs{\CurrentBib}

\bibitem [\protect \citeauthoryear {%
Paschmann%
\ \BBA {} Daly%
}{%
Paschmann%
\ \BBA {} Daly%
}{%
{\protect \APACyear {1998}}%
}]{%
Paschmann1998}
\APACinsertmetastar {%
Paschmann1998}%
\begin{APACrefauthors}%
Paschmann, G.%
\BCBT {}\ \BBA {} Daly, P\BPBI W.%
\end{APACrefauthors}%
\unskip\
\newblock
\APACrefYear{1998}.
\newblock
\APACrefbtitle {{Analysis Methods for Multi-Spacecraft Data}} {{Analysis
  Methods for Multi-Spacecraft Data}}\ (\BVOL~1).
\newblock
\APACaddressPublisher{Keplerlaan}{ESA Publications Division}.
\PrintBackRefs{\CurrentBib}

\bibitem [\protect \citeauthoryear {%
Pierrard%
}{%
Pierrard%
}{%
{\protect \APACyear {1996}}%
}]{%
Pierrard1996}
\APACinsertmetastar {%
Pierrard1996}%
\begin{APACrefauthors}%
Pierrard, V.%
\end{APACrefauthors}%
\unskip\
\newblock
\APACrefYearMonthDay{1996}{}{}.
\newblock
{\BBOQ}\APACrefatitle {{New model of magnetospheric current-voltage
  relationship}} {{New model of magnetospheric current-voltage
  relationship}}.{\BBCQ}
\newblock
\APACjournalVolNumPages{J. Geophys. Res. Sp. Phys.}{101}{A2}{2669--2675}.
\newblock
\begin{APACrefURL} \url{http://dx.doi.org/10.1029/95JA00476} \end{APACrefURL}
\newblock
\begin{APACrefDOI} \doi{10.1029/95JA00476} \end{APACrefDOI}
\PrintBackRefs{\CurrentBib}

\bibitem [\protect \citeauthoryear {%
Pierrard%
, Khazanov%
\BCBL {}\ \BBA {} Lemaire%
}{%
Pierrard%
\ \protect \BOthers {.}}{%
{\protect \APACyear {2007}}%
}]{%
Pierrard2007a}
\APACinsertmetastar {%
Pierrard2007a}%
\begin{APACrefauthors}%
Pierrard, V.%
, Khazanov, G\BPBI V.%
\BCBL {}\ \BBA {} Lemaire, J\BPBI F.%
\end{APACrefauthors}%
\unskip\
\newblock
\APACrefYearMonthDay{2007}{nov}{}.
\newblock
{\BBOQ}\APACrefatitle {{Current--voltage relationship}} {{Current--voltage
  relationship}}.{\BBCQ}
\newblock
\APACjournalVolNumPages{J. Atmos. Solar-Terrestrial Phys.}{69}{16}{2048--2057}.
\newblock
\begin{APACrefURL}
  \url{http://www.sciencedirect.com/science/article/pii/S1364682607002441}
  \end{APACrefURL}
\newblock
\begin{APACrefDOI} \doi{http://dx.doi.org/10.1016/j.jastp.2007.08.005}
  \end{APACrefDOI}
\PrintBackRefs{\CurrentBib}

\bibitem [\protect \citeauthoryear {%
Pierrard%
\ \BBA {} Lazar%
}{%
Pierrard%
\ \BBA {} Lazar%
}{%
{\protect \APACyear {2010}}%
}]{%
Pierrard2010}
\APACinsertmetastar {%
Pierrard2010}%
\begin{APACrefauthors}%
Pierrard, V.%
\BCBT {}\ \BBA {} Lazar, M.%
\end{APACrefauthors}%
\unskip\
\newblock
\APACrefYearMonthDay{2010}{}{}.
\newblock
{\BBOQ}\APACrefatitle {{Kappa Distributions: Theory and Applications in Space
  Plasmas}} {{Kappa Distributions: Theory and Applications in Space
  Plasmas}}.{\BBCQ}
\newblock
\APACjournalVolNumPages{Solar Physics}{267}{1}{153--174}.
\newblock
\begin{APACrefURL} \url{http://dx.doi.org/10.1007/s11207-010-9640-2}
  \end{APACrefURL}
\newblock
\begin{APACrefDOI} \doi{10.1007/s11207-010-9640-2} \end{APACrefDOI}
\PrintBackRefs{\CurrentBib}

\bibitem [\protect \citeauthoryear {%
Press%
, Teukolsky%
, Vetterling%
\BCBL {}\ \BBA {} Flannery%
}{%
Press%
\ \protect \BOthers {.}}{%
{\protect \APACyear {2007}}%
}]{%
Press2007}
\APACinsertmetastar {%
Press2007}%
\begin{APACrefauthors}%
Press, W.%
, Teukolsky, S.%
, Vetterling, W.%
\BCBL {}\ \BBA {} Flannery, B.%
\end{APACrefauthors}%
\unskip\
\newblock
\APACrefYear{2007}.
\newblock
\APACrefbtitle {{Numerical Recipes 3rd Edition: The Art of Scientific
  Computing}} {{Numerical Recipes 3rd Edition: The Art of Scientific
  Computing}}.
\newblock
\APACaddressPublisher{New York}{Cambridge University Press}.
\newblock
\begin{APACrefURL} \url{http://numerical.recipes/} \end{APACrefURL}
\PrintBackRefs{\CurrentBib}

\bibitem [\protect \citeauthoryear {%
\APACciteatitle {{Processes leading to plasma losses into the high-latitude
  atmosphere}}}{%
\APACciteatitle {{Processes leading to plasma losses into the high-latitude
  atmosphere}}}{%
{\protect \APACyear {1999}}%
}]{%
Hultqvist1999}
\APACinsertmetastar {%
Hultqvist1999}%
{\BBOQ}\APACrefatitle {{Processes leading to plasma losses into the
  high-latitude atmosphere}} {{Processes leading to plasma losses into the
  high-latitude atmosphere}}.{\BBCQ}
\newblock
\APACrefYearMonthDay{1999}{}{}.
\newblock
\BIn{} B.~Hultqvist, M.~{\O}ieroset, G.~Paschmann\BCBL {}\ \BBA {} R\BPBI
  A.~Treumann\ (\BEDS), \APACrefbtitle {Magnetos. Plasma Sources Losses Final
  Rep. ISSI Study Proj. Source Loss Process.} {Magnetos. plasma sources losses
  final rep. issi study proj. source loss process.}\ (\BPGS\ 85--135).
\newblock
\APACaddressPublisher{Dordrecht}{Springer Netherlands}.
\newblock
\begin{APACrefURL} \url{http://dx.doi.org/10.1007/978-94-011-4477-3}
  \end{APACrefURL}
\newblock
\begin{APACrefDOI} \doi{10.1007/978-94-011-4477-3} \end{APACrefDOI}
\PrintBackRefs{\CurrentBib}

\bibitem [\protect \citeauthoryear {%
Shiokawa%
\ \protect \BOthers {.}}{%
Shiokawa%
\ \protect \BOthers {.}}{%
{\protect \APACyear {1990}}%
}]{%
Shiokawa1990}
\APACinsertmetastar {%
Shiokawa1990}%
\begin{APACrefauthors}%
Shiokawa, K.%
, Fukunishi, H.%
, Yamagishi, H.%
, Miyaoka, H.%
, Fujii, R.%
\BCBL {}\ \BBA {} Tohyama, F.%
\end{APACrefauthors}%
\unskip\
\newblock
\APACrefYearMonthDay{1990}{aug}{}.
\newblock
{\BBOQ}\APACrefatitle {{Rocket observation of the magnetosphere-ionosphere
  coupling processes in quiet and active arcs}} {{Rocket observation of the
  magnetosphere-ionosphere coupling processes in quiet and active
  arcs}}.{\BBCQ}
\newblock
\APACjournalVolNumPages{Journal of Geophysical Research}{95}{A7}{10679}.
\newblock
\begin{APACrefURL} \url{https://doi.org/10.1029/JA095iA07p10679
  http://doi.wiley.com/10.1029/JA095iA07p10679} \end{APACrefURL}
\newblock
\begin{APACrefDOI} \doi{10.1029/JA095iA07p10679} \end{APACrefDOI}
\PrintBackRefs{\CurrentBib}

\bibitem [\protect \citeauthoryear {%
Stepanova%
\ \BBA {} Antonova%
}{%
Stepanova%
\ \BBA {} Antonova%
}{%
{\protect \APACyear {2015}}%
}]{%
Stepanova2015}
\APACinsertmetastar {%
Stepanova2015}%
\begin{APACrefauthors}%
Stepanova, M.%
\BCBT {}\ \BBA {} Antonova, E\BPBI E.%
\end{APACrefauthors}%
\unskip\
\newblock
\APACrefYearMonthDay{2015}{may}{}.
\newblock
{\BBOQ}\APACrefatitle {{Role of turbulent transport in the evolution of the
  $\kappa$ distribution functions in the plasma sheet}} {{Role of turbulent
  transport in the evolution of the $\kappa$ distribution functions in the
  plasma sheet}}.{\BBCQ}
\newblock
\APACjournalVolNumPages{Journal of Geophysical Research: Space
  Physics}{120}{5}{3702--3714}.
\newblock
\begin{APACrefURL} \url{http://dx.doi.org/10.1002/2014JA020684}
  \end{APACrefURL}
\newblock
\begin{APACrefDOI} \doi{10.1002/2014JA020684} \end{APACrefDOI}
\PrintBackRefs{\CurrentBib}

\bibitem [\protect \citeauthoryear {%
Temerin%
}{%
Temerin%
}{%
{\protect \APACyear {1997}}%
}]{%
Temerin1997}
\APACinsertmetastar {%
Temerin1997}%
\begin{APACrefauthors}%
Temerin, M.%
\end{APACrefauthors}%
\unskip\
\newblock
\APACrefYearMonthDay{1997}{}{}.
\newblock
{\BBOQ}\APACrefatitle {{What do we really know about auroral acceleration?}}
  {{What do we really know about auroral acceleration?}}{\BBCQ}
\newblock
\APACjournalVolNumPages{Adv. Sp. Res.}{20}{415}{1025--1035}.
\newblock
\begin{APACrefURL}
  \url{http://www.sciencedirect.com/science/article/pii/S0273117797005577}
  \end{APACrefURL}
\PrintBackRefs{\CurrentBib}

\bibitem [\protect \citeauthoryear {%
\APACciteatitle {{Theoretical Building Blocks}}}{%
\APACciteatitle {{Theoretical Building Blocks}}}{%
{\protect \APACyear {2003}}%
}]{%
Paschmann2003}
\APACinsertmetastar {%
Paschmann2003}%
{\BBOQ}\APACrefatitle {{Theoretical Building Blocks}} {{Theoretical Building
  Blocks}}.{\BBCQ}
\newblock
\APACrefYearMonthDay{2003}{}{}.
\newblock
\BIn{} G.~Paschmann, S.~Haaland\BCBL {}\ \BBA {} R.~Treumann\ (\BEDS),
  \APACrefbtitle {Auror. Plasma Phys.} {Auror. plasma phys.}\ (\BPGS\ 41--92).
\newblock
\APACaddressPublisher{Dordrecht}{Springer Netherlands}.
\newblock
\begin{APACrefURL} \url{http://dx.doi.org/10.1007/978-94-007-1086-3}
  \end{APACrefURL}
\newblock
\begin{APACrefDOI} \doi{10.1007/978-94-007-1086-3} \end{APACrefDOI}
\PrintBackRefs{\CurrentBib}

\bibitem [\protect \citeauthoryear {%
Treumann%
}{%
Treumann%
}{%
{\protect \APACyear {1999}}%
{\protect \APACexlab {{\protect \BCnt {1}}}}}]{%
Treumann1999a}
\APACinsertmetastar {%
Treumann1999a}%
\begin{APACrefauthors}%
Treumann, R\BPBI A.%
\end{APACrefauthors}%
\unskip\
\newblock
\APACrefYearMonthDay{1999{\protect \BCnt {1}}}{mar}{}.
\newblock
{\BBOQ}\APACrefatitle {{Generalized-Lorentzian Thermodynamics}}
  {{Generalized-Lorentzian Thermodynamics}}.{\BBCQ}
\newblock
\APACjournalVolNumPages{Physica Scripta}{59}{3}{204--214}.
\newblock
\begin{APACrefURL}
  \url{http://stacks.iop.org/1402-4896/59/i=3/a=004?key=crossref.516e5e0dc1941b9941906c543e2b9aed}
  \end{APACrefURL}
\newblock
\begin{APACrefDOI} \doi{10.1238/Physica.Regular.059a00204} \end{APACrefDOI}
\PrintBackRefs{\CurrentBib}

\bibitem [\protect \citeauthoryear {%
Treumann%
}{%
Treumann%
}{%
{\protect \APACyear {1999}}%
{\protect \APACexlab {{\protect \BCnt {2}}}}}]{%
Treumann1999}
\APACinsertmetastar {%
Treumann1999}%
\begin{APACrefauthors}%
Treumann, R\BPBI A.%
\end{APACrefauthors}%
\unskip\
\newblock
\APACrefYearMonthDay{1999{\protect \BCnt {2}}}{jan}{}.
\newblock
{\BBOQ}\APACrefatitle {{Kinetic Theoretical Foundation of Lorentzian
  Statistical Mechanics}} {{Kinetic Theoretical Foundation of Lorentzian
  Statistical Mechanics}}.{\BBCQ}
\newblock
\APACjournalVolNumPages{Physica Scripta}{59}{1}{19--26}.
\newblock
\begin{APACrefURL}
  \url{http://stacks.iop.org/1402-4896/59/i=1/a=003?key=crossref.cf1c459daa57b7e954fc00a86f030530}
  \end{APACrefURL}
\newblock
\begin{APACrefDOI} \doi{10.1238/Physica.Regular.059a00019} \end{APACrefDOI}
\PrintBackRefs{\CurrentBib}

\bibitem [\protect \citeauthoryear {%
Vasyli\={u}nas%
}{%
Vasyli\={u}nas%
}{%
{\protect \APACyear {1968}}%
}]{%
Vasyliunas1968}
\APACinsertmetastar {%
Vasyliunas1968}%
\begin{APACrefauthors}%
Vasyli\={u}nas, V.%
\end{APACrefauthors}%
\unskip\
\newblock
\APACrefYearMonthDay{1968}{}{}.
\newblock
{\BBOQ}\APACrefatitle {A survey of low-energy electrons in the evening sector
  of the magnetosphere with OGO 1 and OGO 3} {A survey of low-energy electrons
  in the evening sector of the magnetosphere with ogo 1 and ogo 3}.{\BBCQ}
\newblock
\APACjournalVolNumPages{Journal of Geophysical Research}{}{}{}.
\newblock
\begin{APACrefDOI} \doi{10.1029/JA073i009p02839} \end{APACrefDOI}
\PrintBackRefs{\CurrentBib}

\bibitem [\protect \citeauthoryear {%
Wing%
\ \BBA {} Newell%
}{%
Wing%
\ \BBA {} Newell%
}{%
{\protect \APACyear {1998}}%
}]{%
Wing1998}
\APACinsertmetastar {%
Wing1998}%
\begin{APACrefauthors}%
Wing, S.%
\BCBT {}\ \BBA {} Newell, P\BPBI T.%
\end{APACrefauthors}%
\unskip\
\newblock
\APACrefYearMonthDay{1998}{apr}{}.
\newblock
{\BBOQ}\APACrefatitle {{Central plasma sheet ion properties as inferred from
  ionospheric observations}} {{Central plasma sheet ion properties as inferred
  from ionospheric observations}}.{\BBCQ}
\newblock
\APACjournalVolNumPages{Journal of Geophysical Research: Space
  Physics}{103}{A4}{6785--6800}.
\newblock
\begin{APACrefURL} \url{http://dx.doi.org/10.1029/97JA02994} \end{APACrefURL}
\newblock
\begin{APACrefDOI} \doi{10.1029/97JA02994} \end{APACrefDOI}
\PrintBackRefs{\CurrentBib}

\bibitem [\protect \citeauthoryear {%
Wygant%
}{%
Wygant%
}{%
{\protect \APACyear {2002}}%
}]{%
Wygant2002}
\APACinsertmetastar {%
Wygant2002}%
\begin{APACrefauthors}%
Wygant, J\BPBI R.%
\end{APACrefauthors}%
\unskip\
\newblock
\APACrefYearMonthDay{2002}{}{}.
\newblock
{\BBOQ}\APACrefatitle {{Evidence for kinetic Alfv{\'{e}}n waves and parallel
  electron energization at 4--6 $R_E$ altitudes in the plasma sheet boundary
  layer}} {{Evidence for kinetic Alfv{\'{e}}n waves and parallel electron
  energization at 4--6 $R_E$ altitudes in the plasma sheet boundary
  layer}}.{\BBCQ}
\newblock
\APACjournalVolNumPages{Journal of Geophysical Research}{107}{A8}{1201}.
\newblock
\begin{APACrefURL}
  \url{http://www.scopus.com/inward/record.url?eid=2-s2.0-84905344954{\&}partnerID=tZOtx3y1}
  \end{APACrefURL}
\newblock
\begin{APACrefDOI} \doi{10.1029/2001JA900113} \end{APACrefDOI}
\PrintBackRefs{\CurrentBib}

\end{thebibliography}




%
%
%
%
%
%
%

  \acknowledgments 

  All observations and measurements made by the FAST spacecraft are available as
  a Level 1 data product through SDT
  (http://sprg.ssl.berkeley.edu/~sdt/SdtReleases.html). We are grateful to Craig
  Markwardt for making publicly available MPFIT
  (http://cow.physics.wisc.edu/~craigm/idl/fitting.html), the Interactive Data
  Language version of the MINPACK-1 fitting routines, which we have used
  extensively in this research. Work at the Birkeland Center for Space Science
  and the University of Bergen was funded by the Research Council of Norway/CoE
  under contract 223252/F50. Work at Dartmouth College was supported by NASA
  Headquarters under NASA grant NNX17AF92G and sub-award W000726838 to NASA
  grant NNX15AL08G. Work at Space Sciences Laboratory was supported by NASA
  grants NNX15AF57G and NNX16AG69G.

\end{document}